\documentclass{article}

\PassOptionsToPackage{numbers, compress}{natbib}

\usepackage{natbib}
\usepackage{multirow}
\usepackage{url}
\usepackage{blindtext}
\usepackage{capt-of}
\usepackage{graphicx}
\usepackage{floatrow}
\newfloatcommand{capbtabbox}{table}[\captop][\FBwidth]
\usepackage{blindtext}
\usepackage{wrapfig}
\usepackage[preprint]{neurips_2021}
\usepackage{array,booktabs}
\usepackage{amsmath}
\usepackage{enumitem}
\setlist[itemize]{leftmargin=0.3in}

\newcolumntype{M}[1]{>{\centering\arraybackslash}m{#1}}

\usepackage[utf8]{inputenc} 
\usepackage[T1]{fontenc}    
\usepackage{hyperref}       
\usepackage{url}            
\usepackage{booktabs}       
\usepackage{amsfonts}       
\usepackage{nicefrac}       
\usepackage{microtype}      
\usepackage{xcolor}         
\usepackage{caption, makecell}

\usepackage{float}
\floatstyle{plaintop}
\restylefloat{table}

\usepackage{authblk}

\def\CC{{C\nolinebreak[4]\hspace{-.05em}\raisebox{.4ex}{\tiny\bf ++}}}

\title{SpeechBrain: A General-Purpose Speech Toolkit}

\author[1,2]{\textbf{Mirco~Ravanelli}}
\author[3,16]{\textbf{Titouan~Parcollet}}
\author[4]{\textbf{Peter~Plantinga}}
\author[5]{\textbf{Aku~Rouhe}}
\author[6]{\textbf{Samuele~Cornell}}
\author[1,7]{\textbf{Loren~Lugosch}}
\author[1]{\textbf{Cem~Subakan}}
\author[8]{\textbf{Nauman~Dawalatabad}}
\author[9]{\textbf{Abdelwahab~Heba}}
\author[1]{\textbf{Jianyuan~Zhong}}
\author[10\thanks{Work conducted while at National Taiwan University.}]{\textbf{Ju-Chieh~Chou}}
\author[11*]{\textbf{Sung-Lin~Yeh}}
\author[12]{\textbf{Szu-Wei~Fu}}
\author[12]{\textbf{Chien-Feng~Liao}}
\author[13\thanks{Work conducted while on an internship at Mila - Quebec AI Institute.}]{\textbf{Elena~Rastorgueva}}
\author[14]{\textbf{François~Grondin}}
\author[14]{\textbf{William~Aris}}
\author[15]{\textbf{Hwidong~Na}}
\author[16]{\textbf{Yan~Gao}}
\author[3,7]{\textbf{Renato~De~Mori}}
\author[1,2]{\textbf{Yoshua~Bengio}}

\affil[1]{Mila - Quebec AI Institute}
\affil[2]{Université de Montréal}

\affil[3]{LIA - Avignon Université}
\affil[4]{Ohio State University}
\affil[5]{Aalto University}
\affil[6]{Università Politecnica delle Marche}
\affil[7]{McGill University}
\affil[8]{Indian Institute of Technology Madras}
\affil[9]{IRIT - Université Paul Sabatier}

\affil[10]{Toyota Technological Institute at Chicago}
\affil[11]{University of Edinburgh}
\affil[12]{Academia Sinica, Taiwan}

\affil[13]{NVIDIA}
\affil[14]{Université de Sherbrooke}
\affil[15]{Samsung-SAIT}
\affil[16]{CaMLSys - University of Cambridge}

\usepackage{listings}
\usepackage{xcolor}

\definecolor{codegreen}{rgb}{0,0.6,0}
\definecolor{codegray}{rgb}{0.5,0.5,0.5}
\definecolor{codepurple}{rgb}{0.58,0,0.82}
\definecolor{backcolour}{rgb}{0.95,0.95,0.92}

\usepackage{inconsolata}
\lstdefinestyle{mystyle}{
    backgroundcolor=\color{backcolour},   
    commentstyle=\color{codegreen},
    keywordstyle=\color{magenta},
    numberstyle=\tiny\color{codegray},
    stringstyle=\color{codepurple},
    basicstyle=\ttfamily\footnotesize,
    breakatwhitespace=false,         
    breaklines=true,                 
    captionpos=b,                    
    keepspaces=true,                 
    numbers=left,                    
    numbersep=5pt,                  
    showspaces=false,                
    showstringspaces=false,
    xleftmargin=14pt,
    framexleftmargin=14pt,
    showtabs=false,                  
    tabsize=2
}

\colorlet{punct}{red!60!black}
\definecolor{background}{HTML}{EEEEEE}
\definecolor{delim}{RGB}{20,105,176}
\colorlet{numb}{magenta!60!black}

\lstdefinelanguage{json}{
    basicstyle=\normalfont\ttfamily,
    numbers=left,
    numberstyle=\tiny\color{codegray},
    xleftmargin=14pt,
    framexleftmargin=14pt,
    stepnumber=1,
    numbersep=8pt,
    showstringspaces=false,
    breaklines=true,
    frame=lines,
    backgroundcolor=\color{backcolour},
    literate=
     *{0}{{{\color{numb}0}}}{1}
      {1}{{{\color{numb}1}}}{1}
      {2}{{{\color{numb}2}}}{1}
      {3}{{{\color{numb}3}}}{1}
      {4}{{{\color{numb}4}}}{1}
      {5}{{{\color{numb}5}}}{1}
      {6}{{{\color{numb}6}}}{1}
      {7}{{{\color{numb}7}}}{1}
      {8}{{{\color{numb}8}}}{1}
      {9}{{{\color{numb}9}}}{1}
      {:}{{{\color{punct}{:}}}}{1}
      {,}{{{\color{punct}{,}}}}{1}
      {\{}{{{\color{delim}{\{}}}}{1}
      {\}}{{{\color{delim}{\}}}}}{1}
      {[}{{{\color{delim}{[}}}}{1}
      {]}{{{\color{delim}{]}}}}{1},
}

\begin{document}
\maketitle

\begin{abstract}
SpeechBrain is an open-source and all-in-one speech toolkit. It is designed to facilitate the research and development of neural speech processing technologies by being simple, flexible, user-friendly, and well-documented. 
This paper describes the core architecture designed to support several tasks of common interest, allowing users to naturally conceive, compare and share novel speech processing pipelines.
SpeechBrain achieves competitive or state-of-the-art performance in a wide range of speech benchmarks. It also provides training recipes, pretrained models, and inference scripts for popular speech datasets, as well as tutorials which allow anyone with basic Python proficiency to familiarize themselves with speech technologies.
\end{abstract}

\section{Introduction}
Open-source toolkits have played a critical role in the development of speech processing technology \cite{Young02, sphinx, julius, rwth, Povey}.
Kaldi \cite{Povey}, for instance, is an established speech recognition framework, which is implemented in \CC{} with recipes built on top of Bash, Perl, and Python scripts. Despite being efficient, its use of \CC{} can make prototyping of new deep learning methods difficult.
With the advent of general-purpose deep learning libraries like TensorFlow \cite{tensorflow} and PyTorch \cite{pytorch}, more flexible speech recognition frameworks have quickly appeared, e.g., DeepSpeech \cite{deepspeech}, RETURNN \cite{returnn}, PyTorch-Kaldi \cite{pytorch_kaldi}, Espresso \cite{espresso}, Lingvo \cite{lingvo}, Fairseq \cite{fairseq}, ESPnet \cite{espnet}, and NeMo \cite{nemo}.

Recently, task-specific libraries have also been released. Examples are Asteroid \cite{asteroid} for speech separation, pyannote \cite{pyannote} and sidekit \cite{sidekit} for speaker diarization, and s3prl \cite{superb} for self-supervised speech representations. 
While excelling at specific tasks, these frameworks have different coding styles, standards, and programming languages,
making it challenging and time-consuming to migrate from one codebase to another.
Moreover, their combination in complex speech processing pipelines poses a challenge for interoperability, as connecting different frameworks might be unnatural and their codebases can interact in unpredictable ways. 

Our experience suggests that having a \textit{single}, \textit{flexible}, \textit{multi-task} toolkit can significantly speed up the development of speech technologies. 
Due to growing interest in end-to-end spoken dialog systems (e.g., virtual assistants), implementing composite pipelines within an integrated toolkit offers many advantages.
A single toolkit, for instance, encourages the exploration of transfer learning and joint training techniques across different tasks \cite{jointraining, Chen2015sep, GaoDDL15, ravanelli_SLT} and enables the creation of fully differentiable graphs where multiple technologies are trained jointly and learn to interact.

Inspired by this vision, we have developed SpeechBrain\footnote{The toolkit website can be found at \url{speechbrain.github.io/}.}, an all-in-one PyTorch-based toolkit designed to facilitate the development, portability, and ease of use of speech processing technologies. 
The name \textit{SpeechBrain} highlights the need for a holistic system capable of performing multiple tasks at once, for example, recognize speech, understanding its content, language, emotions, and speakers. 
Our toolkit is not only intended for speech researchers, but also for the broader machine learning community, enabling users to easily integrate their models into different speech pipelines and compare them with state-of-the-art (SotA) baselines. Our main contributions in this paper are:

\begin{itemize}
    \item The presentation of \textit{SpeechBrain}, with an emphasis on how we designed it to support multiple tasks without sacrificing simplicity, modularity, or flexibility.
    \item The implementation and experimental validation of both recent and long-established speech processing models with SotA or competitive performance on a variety of tasks (cf. Table \ref{tab:tasks}).
\end{itemize}
More broadly, we believe the SpeechBrain toolkit has the potential to significantly accelerate research and innovation in the field of speech processing and deep learning.

\begin{table}[t!]
\caption{List of speech tasks and corpora that are currently supported by SpeechBrain.}
  \centering
  \small
    \begin{tabular}{p{27mm}p{36mm}p{35mm}p{25mm}} \toprule
        \textbf{Task}   & \textbf{Description}  & \textbf{Techniques} & \textbf{Datasets} \\ \midrule
   Speech recognition & \textit{Speech-to-text.} & CTC \cite{CTC} \newline Transducers \cite{graves_transducer} \newline CTC+Attention \cite{hybrid_ctc_att} \newline Shallow fusion \cite{shallow_fusion} & LibriSpeech \cite{librispeech} \newline Common Voice \cite{commonvoice:2020} \newline AISHELL \cite{aishell_2017} \newline TIMIT \cite{timit}\\ \midrule 
 Speaker recognition & \textit{Speaker verification/ID.} & X-vectors \cite{xvector} \newline ECAPA-TDNN \cite{ecapa}  & VoxCeleb1 \cite{voxceleb} \newline VoxCeleb2 \cite{voxceleb2}\\
\hline
Speaker diarization & \textit{Detect who spoke when.} & Spectral Clustering \cite{spec_tutorial} \newline Neural embeddings \cite{ecapa_diarization} & AMI corpus \cite{ami-corpus} \\
\hline
Speech enhancement & \textit{Noisy to clean speech.} & MetricGAN+ \cite{fu2021metricgan+} \newline Mimic Loss \cite{bagchi2018spectral} & VoiceBank \cite{voicebank} \newline DNS \cite{dns_challenge} \\
\hline
Speech separation & \textit{Separate overlapped speech.} & ConvTasNet\cite{convtasnet} \newline DualPath RNNs \cite{luo2020dualpath} \newline SepFormer \cite{sepformer}  & WSJ-mix \cite{deepclustering} \newline WHAM \cite{wham} \newline WHAMR \cite{whamr} \newline
LibriMix \cite{librimix}\\
\hline
Spoken language \newline understanding & \textit{Speech to intent/slots.} & Decoupled \cite{timers-and-such} \newline Multistage \cite{Haghani2018} \newline Direct \cite{Serdyuk2018} & TAS \cite{timers-and-such} \newline SLURP \cite{slurp} \newline FSC \cite{fluent}\\
\hline
Multi-microphone \newline processing & \textit{Combining input signals.} & Delay-and-sum \newline
MVDR~\cite{habets2009new} \newline GEV~\cite{heymann2016neural} \newline
GCC-PHAT~\cite{KnappCarter} \newline SRP-PHAT~\cite{cobos2010modified} \newline MUSIC~\cite{schmidt1986multiple} & Dataset-Independent \\ \bottomrule
\end{tabular}
\label{tab:tasks}
\end{table}

\section{Related Work}
A few other toolkits support multiple speech tasks.
Of these, the ones we consider most related to SpeechBrain are Fairseq \cite{fairseq}, NeMo \cite{nemo}, and ESPnet \cite{espnet}.
Fairseq is developed by Facebook to support sequence-to-sequence processing. It includes models such as ConvS2S \cite{convs2s}, transformers \cite{transformers}, and wav2vec \cite{wav2vect}.
However, speech processing encompasses several paradigms outside of sequence-to-sequence modeling.
SpeechBrain also supports regression tasks (e.g., speech enhancement, separation), classification tasks (e.g., speaker recognition), clustering (e.g., diarization), and even signal processing techniques (e.g., multi-microphone combination). 

NeMo is a toolkit for conversational AI developed by NVIDIA, which provides useful neural modules for many speech processing tasks, including speech recognition, speaker diarization, voice-activity detection and text-to-speech. Due to its industrial orientation, NeMo offers efficient ready-to-use models, such as Jasper \cite{jasper}, QuartzNet \cite{quartznet}, and Citrinet \cite{citrinet}.
SpeechBrain also provides several ready-to-use models, but focuses more heavily on research and education by providing a wide variety of baselines, models, and recipes that users can easily inspect and modify in the experiments.

ESPnet, in its current form, is the closest toolkit to SpeechBrain. Both are academically driven and support numerous speech tasks. 
ESPnet started as an end-to-end speech recognition library and progressively grew to support different tasks.
By contrast, we designed SpeechBrain to address a wide variety of tasks from the outset. This means that combining technologies and developing recipes for new tasks is extremely simple.

\section{Design Principles}
\label{sec:principles}
Beyond the multi-task vision highlighted in the introduction,  we developed SpeechBrain with the following design principles in mind:
    
\textbf{Accessibility}: 
SpeechBrain is designed to be easily understandable by a large user base, including early students and practitioners.
Therefore, we devoted considerable effort to develop intuitive modules that are easy to interconnect with each other.
One remarkable peculiarity of SpeechBrain is that it serves educational purposes as well.  
We thus have written extensive documentation and tutorials with Google Colab to help newcomers become more familiar with speech technologies.
Prior work has shown code snippets aid in adopting a codebase~\cite{fairbanks-2006-design-fragments}. Motivated by this, SpeechBrain provides runnable code snippets in docstrings (documenting interaction at the granular level), tutorial notebooks (explaining single topics), and template files (describing full experiments on different tasks).
To make our toolkit as accessible as possible, we have released it under a very permissive license (Apache 2.0).
    
\textbf{Ease of use}: 
SpeechBrain employs a simple software stack (i.e., Python $\rightarrow$ PyTorch $\rightarrow$ SpeechBrain) to avoid dealing with too many levels of abstractions.
It is developed on top of PyTorch directly, without an external API. PyTorch-compatible code works in our toolkit without any further modification.
SpeechBrain has a minimal list of external dependencies that are all installable via PyPI. The installation process simply requires running the command \colorbox{backcolour}{\lstinline{pip install speechbrain}} and is done within a few minutes. The code is Pythonic and maximizes the use of PyTorch routines.

\textbf{Replicability}: SpeechBrain promotes open and transparent science. We trained most of our models with publicly available data. This way, our results can be easily replicated by the community. 
Several pre-trained models, which only require a few lines of code to use, are distributed via Hugging Face \cite{hugging}.
Besides sharing the code and the trained models, we also share the whole experiment folder, which contains all the needed details (e.g., logs) to reproduce our results. 

\begin{figure}[t!]
\begin{floatrow}
\ffigbox{%
\centering
\includegraphics[trim={0cm 0 0 0.0},clip,width=6.0cm]{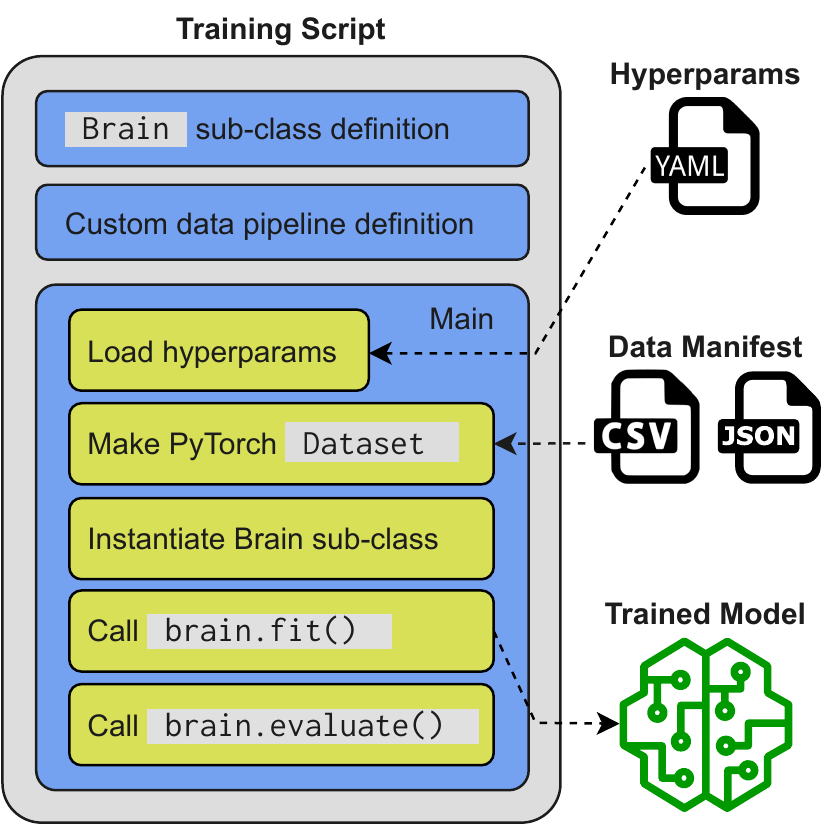}
}{%
  \caption{An overview of a basic training script.}
  \label{fig:arch}
}
\ffigbox{%
    \includegraphics[width=6cm]{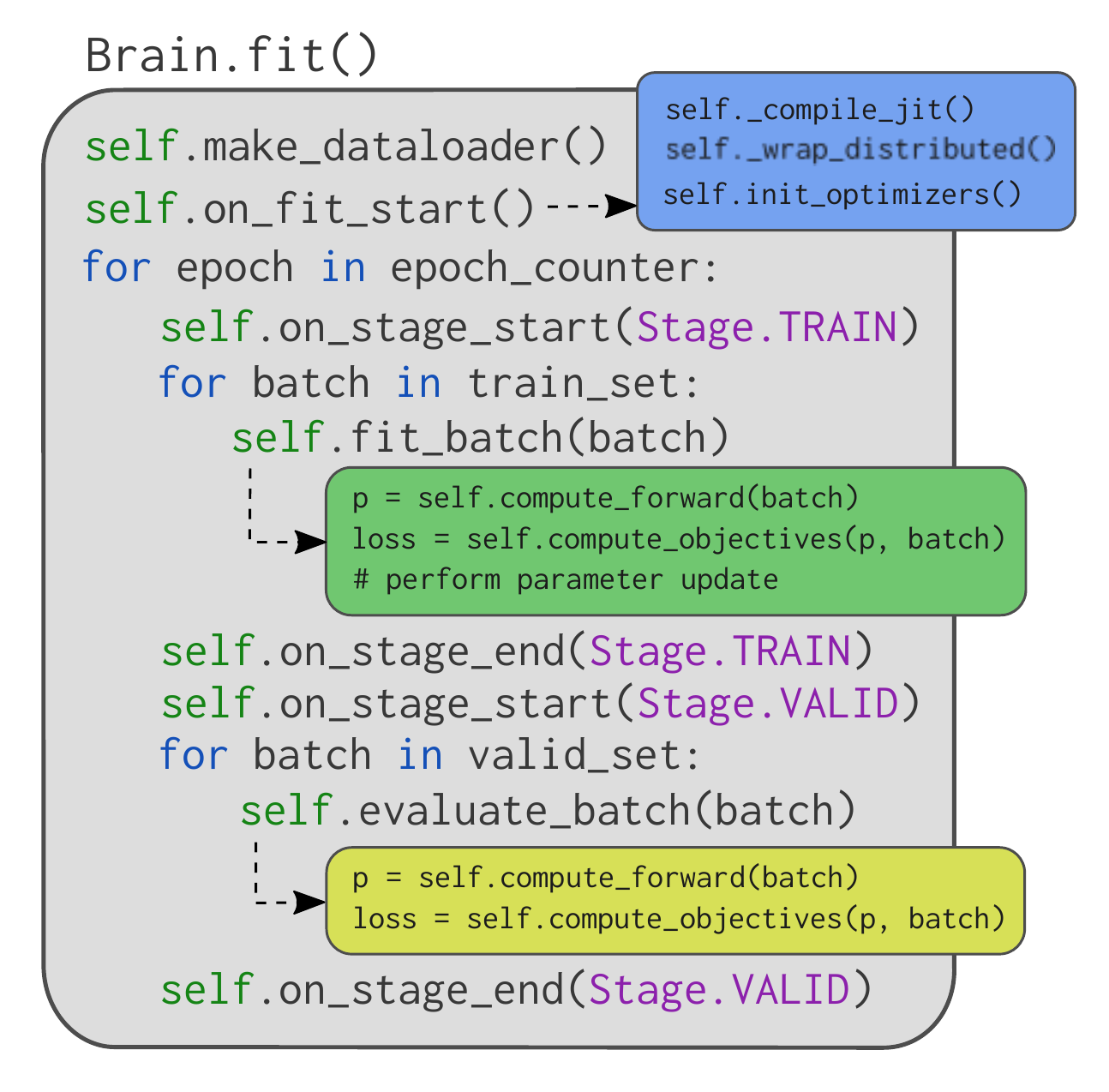}
}{%
  \caption{Illustration of \lstinline{Brain.fit()}.}
  \label{fig:brain_illustration}
}
\end{floatrow}
\end{figure}

\section{Architecture}
\label{sec:sb_arch}
From an architectural standpoint, SpeechBrain sits in between a library and a framework. Where libraries require users to manage dataflow by calling library-defined functionality, frameworks primarily define a custom lifecycle in which user-defined functionalities are invoked in specific places (\textit{inversion of control}).
Most code in SpeechBrain follows a library-style collection of modular and standalone building blocks, including practical routines for data loading, decoding, signal processing, and other convenient utilities.
However the central \colorbox{backcolour}{\lstinline{Brain}} class (see \S~\ref{sec:brain_api}), uses inversion of control to define a general training loop.
Therefore, SpeechBrain is most accurately described as a \textit{toolkit}.
As shown in Figure~\ref{fig:arch}, the code for training a model is contained within a single Python script. Training begins by calling the script with a set of hyperparameters: \colorbox{backcolour}{\lstinline{python train.py hparams.yaml}}. These hyperparameters, declared in human-readable YAML format, contain the location of one or more data manifest files using either CSV or JSON formats (see Appendix \ref{app:arch_det}).
Unlike many other toolkits, SpeechBrain orchestrates experiments in Python directly, without relying on external Bash scripts. This allows code for data loading, modeling, optimization, and evaluation to interact naturally. Moreover, the training script exposes the computations likely to be changed most frequently (e.g., forward computations, data transformations, etc.), making them easy to access and modify.
SpeechBrain treats the user's code as a first-class citizen: all PyTorch-compatible code written by the user is treated the same as SpeechBrain code. In the following sub-sections, we explore the anatomy of a training script in more detail.

\subsection{Hyperparameters}
The model hyperparameters, in conjunction with the training script, regulate various properties of the pipeline such as model architecture, training, and decoding. 
SpeechBrain relies on an extended version of YAML called \textit{HyperPyYAML},
as shown in the following excerpt:

\begin{lstlisting}[language=json, firstnumber=1, basicstyle=\footnotesize\ttfamily, frame=lines, caption=An excerpt of a YAML file for hyperparameter specification.]
dropout: 0.2
features: !new:speechbrain.lobes.features.MFCC
    n_mels: 40
    left_frames: 5
    right_frames: 5

model: !new:torch.nn.LSTM
   input_size: 440
   hidden_size: 256
   num_layers: 4
   dropout:  !ref <dropout>
   bidirectional: True
\end{lstlisting}

HyperPyYAML is not just an ordinary list of hyperparameters, but allows a complex hyperparameter specification that defines objects along with their corresponding arguments. There is always an explicit reference between the hyperparameter declarations and any object using them, making the code more interpretable and simpler to debug. Overriding the contents of the YAML file (e.g., for hyperparameter search) can also be done easily by passing command-line arguments:

\begin{lstlisting}[language=bash, basicstyle=\footnotesize\ttfamily]
$ python train.py hparams.yaml --learning_rate=0.1 --dropout=0.5
\end{lstlisting}

SpeechBrain initializes the classes automatically when reading the YAML file, thus eliminating boilerplate initialization code from the training script. HyperPyYAML is a general tool for specifying hyperparameters. To enable modular reusuability, we have released it as a separate repository on PyPI\footnote{\url{github.com/speechbrain/HyperPyYAML}}.

\subsection{Data loading}\label{sec:data_loading}
SpeechBrain complements standard PyTorch data loading by addressing the typical challenges that occur when working with speech, such as
handling variable-length sequences, large datasets, and complex data transformation pipelines.
Our \colorbox{backcolour}{\lstinline{DynamicItemDataset}} inherits from \colorbox{backcolour}{\lstinline{torch.utils.data.Dataset}} and 
creates a dataset-interface based on a data-manifest file.
The data-manifest file contains \textit{static items}, such as filepaths or speaker labels. Then, \textit{dynamic items} 
provide transformations based on the existing items (static or dynamic), as shown in the following example:

\begin{lstlisting}[language=Python, frame=lines, basicstyle=\footnotesize\ttfamily, caption={An example of a custom data pipeline.}, label={lst:data_pip}]
@speechbrain.utils.data_pipeline.takes("file_path")
@speechbrain.utils.data_pipeline.provides("signal")
def audio_pipeline(file_path):
  return speechbrain.dataio.read_audio(file_path)
\end{lstlisting}
This function takes an audio file path (a static item) and
reads it as a tensor called "signal" (a dynamic item). Any library for reading audio file can be used here, including torch.audio\footnote{\url{https://github.com/pytorch/audio}}. 
The evaluation order of the items is determined by a dependency graph.
Users can define operations such as reading and augmenting an audio file, encoding a text label into an integer, basic text processing, etc. 
The dynamic items are defined in the training script and are thus directly customizable by the users.
Moreover, by leveraging the PyTorch \colorbox{backcolour}{\lstinline{DataLoader}} class, these data pipelines are automatically applied in parallel across different workers. 

\subsection{Batching}
Speech sequences for a given dataset typically vary in length and require zero-padding to create equal-length batches. This tends to add some complication during the training process. 
First, the length of each sentence within each batch must be tracked so we can later remove zero-padded elements from computations like normalization, statistical pooling, losses, etc.
Another issue that arises is how to avoid wasting computational resources processing zero-padded elements.
 
One approach to mitigate this issue is to sort data by sequence length before batching, which minimizes zero-padding but sacrifices randomness in the batch creation process. A more sophisticated approach is to apply dynamic batching \cite{batching,morishita-etal-2017-empirical}, where sentences are clustered by length and sampled within the same cluster, a trade-off between random and sorted batching. This allows the batch size to be dynamically changed according to sentence length, leading to improved efficiency and better management of available GPU memory.
All the aforementioned batching strategies are supported by SpeechBrain, allowing users to choose the approach that meets their specific needs.

\subsection{The \colorbox{backcolour}{\lstinline{Brain}} class}\label{sec:brain_api}
SpeechBrain implements a general training loop in
the \colorbox{backcolour}{\lstinline{Brain}} class. The \colorbox{backcolour}{\lstinline{Brain.fit()}} method is inspired by similar methods in libraries such as Scikit-learn \cite{pedregosa2011scikit}, Scipy~\cite{virtanen2020scipy}, Keras~\cite{gulli2017deep}, fastai~\cite{howard2020fastai}, and PyTorch Lightning~\cite{falcon2019pytorch}. Figure~\ref{fig:brain_illustration} illustrates the basic components of the \colorbox{backcolour}{\lstinline{Brain.fit()}} method. The following is a simple demonstration:

\begin{lstlisting}[language=Python, frame=lines, basicstyle=\footnotesize\ttfamily, caption=Training a simple model with SpeechBrain using the Brain class.]
import torch, speechbrain

class SimpleBrain(speechbrain.Brain):
  def compute_forward(self, batch, stage):
    return self.modules.model(batch["input"])
  def compute_objectives(self, predictions, batch, stage):
    return torch.nn.functional.l1_loss(predictions, batch["target"])

modules = {"model": torch.nn.Linear(in_features=10, out_features=10)}
brain = SimpleBrain(modules, lambda x: torch.optim.SGD(x, 0.1))
data = [{"input": rand(10, 10), "target": rand(10, 10)}]
brain.fit(epoch_counter=range(15), train_set=data)
\end{lstlisting}

With only about ten lines of code, we can train a neural model. Repetitive boilerplate, such as setting \colorbox{backcolour}{\lstinline{train()}} and \colorbox{backcolour}{\lstinline{eval()}} flags, putting the models on the specified device, and computing gradients are handled by the \colorbox{backcolour}{\lstinline{Brain}} class. Users can override any step of the process, allowing the definition of more complicated (e.g., GAN \cite{gan}) training procedures. The \colorbox{backcolour}{\lstinline{Brain}} class also handles validation, learning rate scheduling, and fault-tolerant model checkpointing, so that training can resume where it left off if execution is interrupted (e.g., by preemption on a cluster). Further details about the \colorbox{backcolour}{\lstinline{Brain}} API are provided in \S~\ref{sec:brain_api_details}. 

\subsection{Other features}
Beyond the functionalities mentioned in the previous sections, additional features include:

\textbf{Multi-GPU training}: SpeechBrain supports both \colorbox{backcolour}{\lstinline{DataParallel}} and \colorbox{backcolour}{\lstinline{DistributedDataParallel}} modules, allowing the use of GPUs on the same and different machines. Automatic mixed-precision can be enabled by setting a single flag to reduce the memory footprint of the models. Moreover, the library supports PyTorch's Just-In-Time (JIT) compiler for native compilation.

\textbf{Large-scale experiments}:
SpeechBrain extends WebDataset\footnote{\url{https://github.com/webdataset/webdataset}} with on-the-fly dynamic batching and bucketing. This enables efficient batching in sequential shard-based data reading, which is necessary for processing large corpora on network filesystems.

\textbf{On-the-fly feature generation}: Rather than serializing intermediate features to disk, SpeechBrain loads raw waveforms and supports a wide variety of efficient streaming operations for audio processing. Standard features like the Short-Term Fourier Transform (STFT) and Mel-filterbanks are computed at training time, allowing differentiation and waveform-level augmentation \cite{Park2019}. Many recipes include on-the-fly augmentations such as adding noise, time warping, or feature masking.

\begin{figure}[t!]
\CenterFloatBoxes
\begin{floatrow}
\capbtabbox{%
    \begin{tabular}{p{18mm}c c c} \toprule
        \textbf{Technique}    & \textbf{\# Params} & \textbf{Dev} & \textbf{Test} \\ \midrule
   CTC  & 10 M & 12.34 & 14.15\\ 
   Transducer  & 10 M & 12.66 & 14.12\\ 
   CTC+Att  & 10 M & 12.74 & 13.83\\ 
   CTC+Att+SSL & 318 M & \textbf{7.11} & \textbf{8.04}\\ \bottomrule
    \end{tabular}
    
}{%
  \caption{Phoneme Error Rate (PER\%) achieved with SpeechBrain on TIMIT using different speech recognizers.}%
  \label{tab:timit}
}
\ffigbox{%
  \includegraphics[trim={0.7cm 0.7cm 1.6cm 1.4cm},clip,scale=0.43]{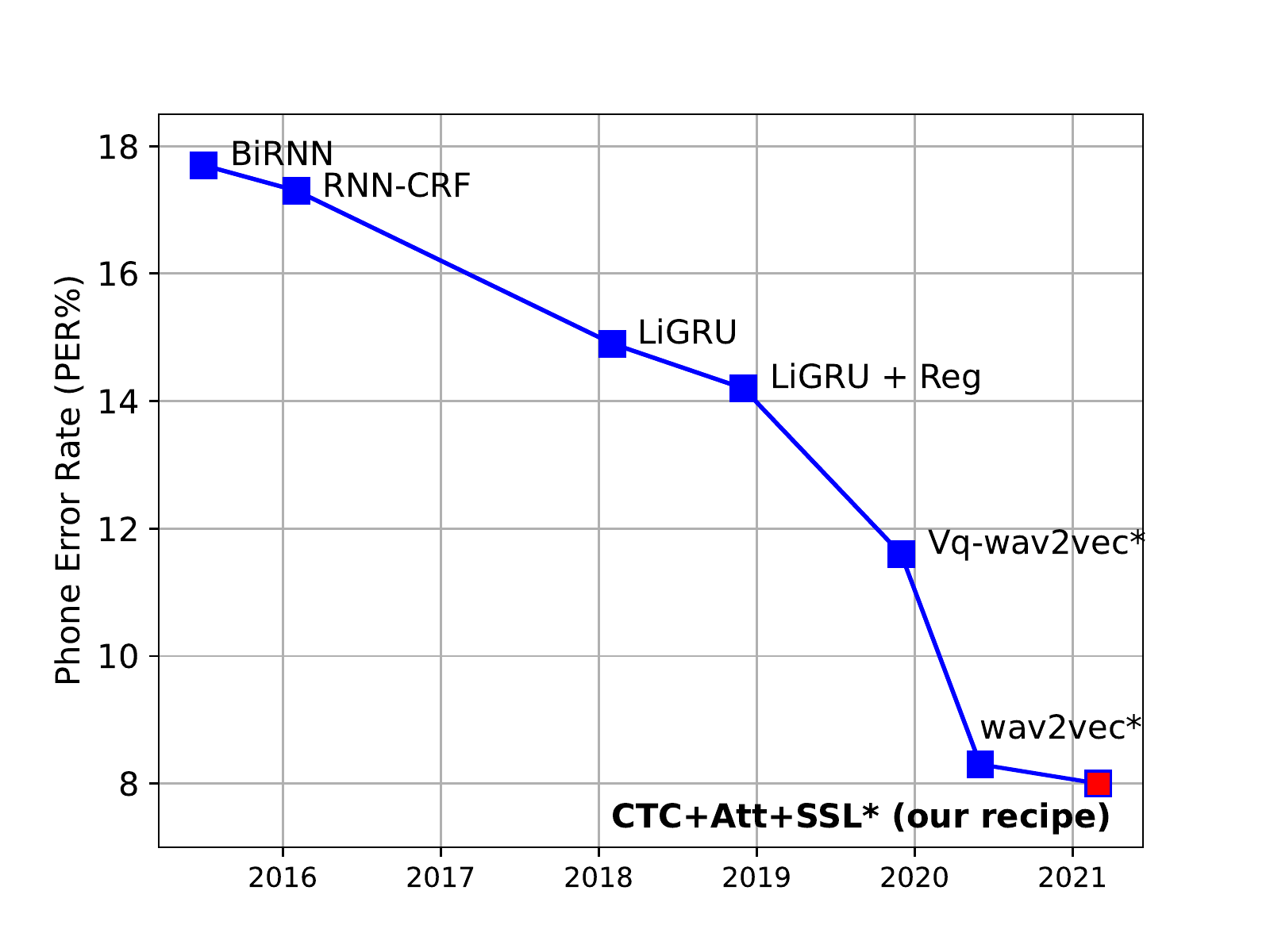}
}{%
  \caption{Evolution of the SotA performance for TIMIT. Entries marked with * use extra unlabelled data from the Libri-Light dataset. Source: \url{https://paperswithcode.com}.}
  \label{fig:timit_soa}
}
\end{floatrow}
\end{figure}

\section{Results}
\label{sec:results}
This section describes use cases of SpeechBrain, highlighting the techniques implemented and the corresponding performance. For more details on datasets, models, and experimental settings, please refer to the appendix (\S~\ref{sec:details}). 

\subsection{Speech recognition}
The toolkit supports common techniques for end-to-end speech recognition with different levels of complexity. The simplest system employs an encoder trained with Connectionist Temporal Classification (CTC) \cite{graves_ctc}. An alternative model is the Transducer \cite{graves_transducer}, which augment CTC with an autoregressive component and a prediction network. The toolkit supports attention-based encoder-decoder architectures as well \cite{hybrid_ctc_att}. In particular, CTC+Att systems rely on an encoder-decoder architecture with an additional CTC loss applied on the top of the encoder. SpeechBrain is designed such that users can easily plug in any encoder and decoder modules into the speech recognition pipeline. For instance, we implemented an effective CRDNN encoder, which combines convolutional, recurrent (e.g., LSTM \cite{lstm}, GRU \cite{gru2}, Light GRU \cite{li_gru}), and fully connected neural networks. As an alternative, users can plug in one of the transformers that we have made available.
Pre-training based on self-supervised learning (SSL) with wav2vec 2.0 \cite{wav2vect} is supported.

We also implemented an efficient GPU-based beam search that combines the acoustic and the language information to retrieve the final sequence of words. The training scripts for language models and tokenizers (using SentencePiece \cite{sentencepiece}) are provided as well.
In the following, we report the performance achieved with SpeechBrain recipes on some popular speech benchmarks.

\subsubsection{TIMIT}
TIMIT \cite{timit} is a small speech dataset with expert-labeled phone sequences. Table \ref{tab:timit} reports the Phone Error Rate (PER) achieved with the aforementioned techniques. All systems use a CRDNN encoder, except for the CTC+Att+SSL one which uses a pre-trained wav2vec 2.0 encoder \cite{wav2vect}.
We report the average performance out of five runs with different random seeds. The standard deviation ranges between 0.15\% and 0.2\% in all the models.

CTC and Transducers provide similar results, while the combination of CTC and attention (CTC+Att) reaches the best performance. The results achieved by our best model (PER 13.8\%) is SotA for TIMIT performance with no extra data. 
A considerable improvement in PER is observed when Light-GRUs \cite{li_gru} are used instead of GRUs \cite{gru2} or LSTMs \cite{lstm} in the CRDNN encoder.
We also observe a  performance boost when using self-supervised pre-training with the wav2vec model trained on unlabelled data from the Libri-Light dataset (CTC+Att+SSL) \cite{librilight}. Our result with this Libri-Light self-supervised pre-training (PER of 8.04\%) slightly outperforms the previous SotA performance with the same pre-training data (PER of 8.30\%), as shown in Figure~\ref{fig:timit_soa}.

\subsubsection{LibriSpeech}\label{librispeech_section}
LibriSpeech \cite{librispeech} is a popular speech recognition benchmark derived from audiobooks.
Table \ref{tab:librispeech_asr} reports the results achieved with different SpeechBrain recipes on this dataset. 
\begin{table}[h!]
    \centering
     \caption{Word Error Rate (WER$\%$) achieved on LibriSpeech with SpeechBrain.}%
    \begin{tabular}{p{20mm}p{16mm}p{14mm}p{14mm} c c} \toprule
        \textbf{Technique}   & \textbf{Encoder} & \textbf{Decoder} & \textbf{\# Params} & \textbf{\lstinline{test-clean}} & \textbf{\lstinline{test-other}} \\ \midrule
   CTC+Att & CRDNN & GRU & 230 M & 2.91 & 8.07 \\ 
   CTC+Att & Transformer & GRU & 161 M & 2.46 & 5.77 \\ 
   \bottomrule 
    \end{tabular}    
  \label{tab:librispeech_asr}
\end{table}

Our best system is a transformer \cite{transformers} combined with a convolutional front-end based on ContextNet \cite{contextnet}. The autoregressive decoder estimates 5k subword tokens derived from running byte-pair encoding on top of the training transcriptions \cite{sentencepiece}. A transformer-based LM is trained on the LibriSpeech text corpus and used within the beam search to rescore partial hypotheses. The best WER that we have achieved on the \lstinline{test-clean} dataset is 2.46\%. This performance is comparable with the results reached in the literature when using transformers without additional data \cite{transformers_asr}. As one can note, the LibriSpeech task is almost perfectly solved by modern speech recognizers. 
We thus focus on more realistic tasks as well, as suggested in some recent works \cite{szymanski-etal-2020-wer, rethinking}. See the appendix (\S~\ref{app:comp_toolkits}) for a more detailed comparison with other toolkits on LibriSpeech and other tasks.

\begin{table}[t!]
\begin{floatrow}
\capbtabbox{%
    \begin{tabular}{p{45mm} c} \toprule
        \textbf{Technique} & \textbf{EER(\%)} \\ \midrule
   VoxCeleb2 baseline \cite{voxceleb2}  &  3.95 \\ 
   Kaldi x-vector \cite{xvector}  &  3.10 \\ 
   ResNET-50 \cite{FITPUB12224}  &  1.19 \\ 
   ECAPA (original paper) \cite{ecapa}  &  0.87 \\ 
     
   \midrule
   SpeechBrain x-vector + PLDA  &  3.20 \\ 
   SpeechBrain ECAPA &  0.81 \\ 
   SpeechBrain ECAPA (vox1+2) &  \textbf{0.69} \\ \bottomrule
    \end{tabular}
    
}{%
  \caption{Equal Error Rate (EER $\%$) achieved on VoxCeleb1 - Cleaned dataset.}%
  \label{tab:speaker_verification}
}
\capbtabbox{%
\resizebox{.45\textwidth}{!}{
    \begin{tabular}{p{33mm}p{10mm}p{10mm}} \toprule
        \textbf{Technique} & \textbf{Known \# spks} & \textbf{Estim. \# spks} \\ \midrule
   MCGAN~\cite{pal21-meta}  &  4.49 &  5.38 \\
   ClusterGAN~\cite{pal21-meta}  &  3.91 &  8.16 \\
   xvector+MCGAN~\cite{pal21-meta}  &  4.23 &  4.92 \\
   xvector+ClusterGAN~\cite{pal21-meta}  &  3.60 &  \textbf{2.87} \\
     VBx (ResNet101) \cite{landini2020VBX}    &  --- &  4.58 \\
   \midrule
   SpeechBrain ECAPA &  \textbf{2.82} &  3.01 \\ \bottomrule  
    \end{tabular}
    }
}{%
  \caption{Diarization Error Rate (DER$\%$) on the eval set of the AMI corpus.}%
  \label{tab:ami_der}
}
\end{floatrow}
\end{table}

\subsubsection{Common Voice}
The Common Voice corpus \cite{commonvoice:2020} is a multilingual open-source collection of transcribed speech based on crowdsourcing data collection. 
CommonVoice is challenging due to significant accented speech, hesitations, presence of foreign words, noise, reverberation,  and other recording artifacts.

Table \ref{tab:commonvoice_asr} reports the results obtained on four different languages.
No language models are trained for this task. The best results are obtained with a wav2vec 2.0 encoder pre-trained with 100k hours of multilingual data from the VoxPopuli dataset \cite{wang2021voxpopuli}. Except for English, the best systems use a GRU decoder on the top of the pre-trained transformer.
CommonVoice is a newer dataset, and there have been relatively few systems evaluated on it. To the best of our knowledge, however, our results are SotA for these languages.

\begin{table}[h!]
\caption{Word Error Rate (WER$\%$) achieved with Common Voice Corpus 6.1 using SpeechBrain on the English (En), French (Fr), Italian (It), and Kinyarwanda (Kw) subsets.}
    \centering
    \begin{tabular}{p{25mm} c c c c c c c c} \toprule
        \textbf{Technique}  & \textbf{Encoder} &\textbf{Decoder} & \textbf{\# Params} & \textbf{En} & \textbf{Fr} & \textbf{It} & \textbf{Kw}\\ \midrule
   CTC+Att  & CRDNN & GRU & 148M & 24.89 & 17.70 & 16.61 & 24.27 \\ 
   CTC+SSL  & Transformer & - & 320M & \textbf{15.58} & 14.44 & 10.93 & 23.12 \\ 
   CTC+Att+SSL & Transformer  & GRU & 330M & 15.69 & \textbf{13.34} & \textbf{9.86} & \textbf{18.91} \\ 
   \bottomrule
    \end{tabular}
  \label{tab:commonvoice_asr}
\end{table}

\subsection{Speaker recognition and diarization}
SpeechBrain implements the functionalities needed to support speaker recognition and speaker diarization. It supports popular embeddings 
derived from Time Delay Neural Networks (TDNNs) \cite{Lang+Hinton88,waibel}, such as x-vectors \cite{xvector} and the recent ECAPA-TDNN embeddings \cite{ecapa}.   
Furthermore, SpeechBrain provides traditional Probabilistic Linear Discriminant Analysis (PLDA) for speaker discrimination \cite{Kenny-plda,GarciaRomero2011AnalysisOI}. 

Table \ref{tab:speaker_verification} reports the performance achieved on a speaker verification task with models trained on VoxCeleb2 \cite{voxceleb2} and tested on VoxCeleb1-clean \cite{voxceleb}.
The best model for speaker embeddings available in SpeechBrain is the ECAPA-TDNN, which matches the performance achieved in the original paper \cite{ecapa}. This model outperforms both the x-vectors \cite{xvector} and the ResNet-34 \cite{FITPUB12224} by a large margin. To the best of our knowledge, the EER reached so far by SpeechBrain on VoxCeleb is the best so far reached by an open-source toolkit.

Table \ref{tab:ami_der} reports the performance achieved on speaker diarization with the AMI meeting corpus \cite{ami-corpus} when using the embeddings available in SpeechBrain. 
In this case, the embeddings are clustered with spectral clustering to assign a relative speaker label to each segment of the recording \cite{ecapa_diarization}.
The results shown are obtained on the official Full-ASR split of the AMI corpus while keeping 0.25 sec of forgiveness collar.
The best diarization system available in SpeechBrain outperforms recent approaches based on meta-learning (MCGAN/ClusterGAN) \cite{pal21-meta}, and Variational Bayes (VBx) \cite{landini2020VBX} when the number of speakers is known (e.g., in a meeting). We have also obtained competitive results when the number of speakers is unknown.

\subsection{Speech enhancement and separation}

SpeechBrain supports speech enhancement models with different input features (e.g., spectral and waveform domain) and training losses (e.g., L1, MSE, and STOI). In addition, it supports a variety of more sophisticated multi-model training techniques such as Mimic Loss~\cite{bagchi2018spectral} and MetricGAN+~\cite{fu2021metricgan+}.
\begin{figure}
\CenterFloatBoxes
\begin{floatrow}
\capbtabbox{%
    \setlength{\tabcolsep}{4pt}
    \begin{tabular}{lccc} \toprule
        \textbf{Technique} & \textbf{\# Params} & \textbf{PESQ} & \textbf{COVL}  \\ \midrule
    \begin{tabular}{@{}l@{}}Facebook \\ DEMUCS~\cite{defossez2020real}\end{tabular} & 60.8 M &  3.07 & 3.63 \\
    \midrule
    \begin{tabular}{@{}l@{}}SpeechBrain \\ Mimic Loss\end{tabular} & 22.3 M &  3.05 & \textbf{3.74} \\[10pt]
    \begin{tabular}{@{}l@{}}SpeechBrain \\ MetricGAN+\end{tabular} & 1.9 M &  \textbf{3.15} & 3.62 \\
    \bottomrule
    \end{tabular}
}{%
  \caption{Speech enhancement performance on VoiceBank-DEMAND.}%
  \label{tab:voicebank-compare}%
}
\ffigbox{%
  \includegraphics[trim={0.4cm 0.7cm 1.6cm 1.0cm},clip,scale=0.43]{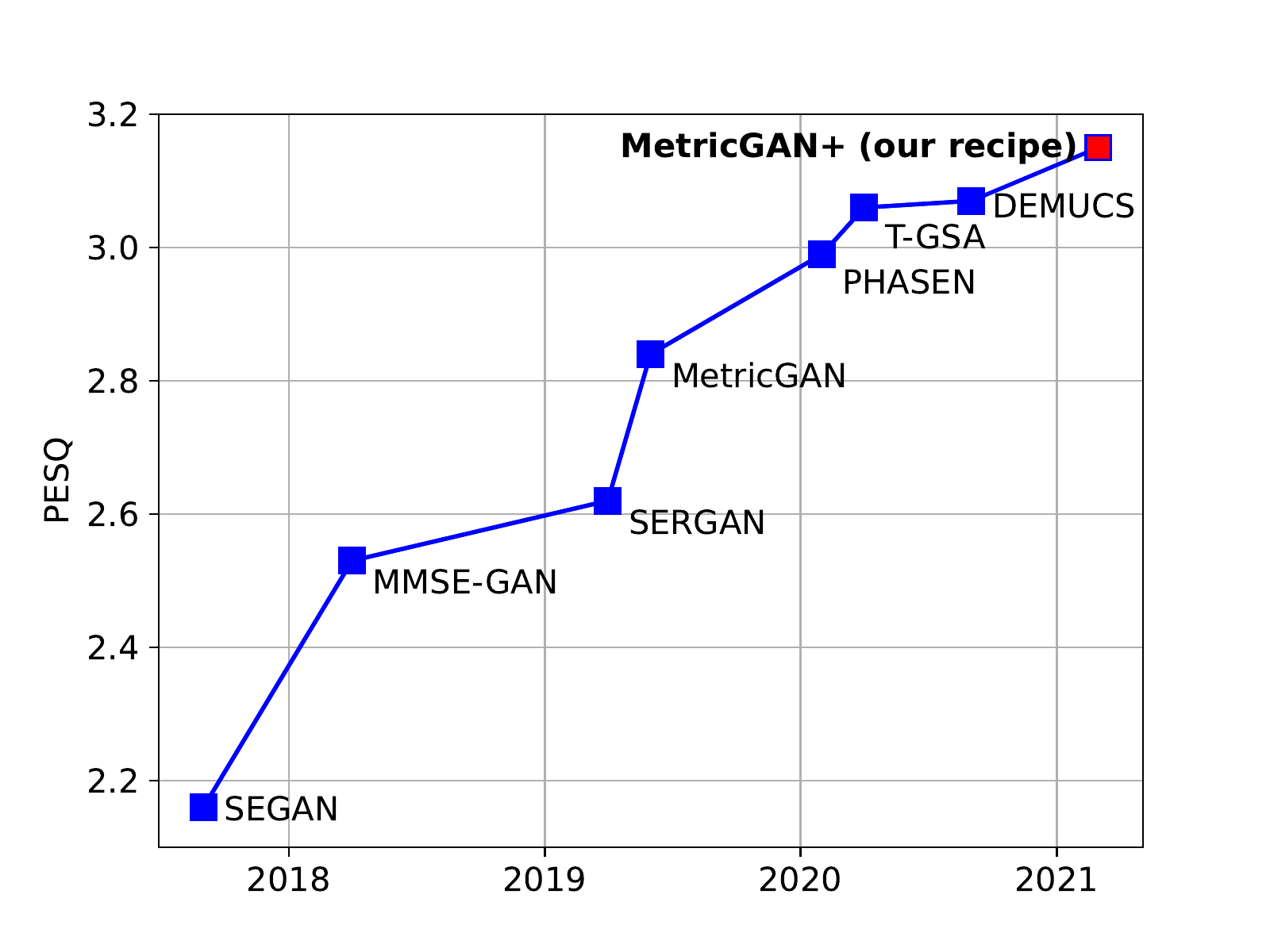}
}{%
  \caption{Evolution of the speech enhancement performance (PESQ) for Voicebank-DEMAND.}%
  \label{fig:voicebank-evolution}%
}
\end{floatrow}
\end{figure}

In Table~\ref{tab:voicebank-compare} we compare the best enhancement systems available in SpeechBrain against the SotA DEMUCS model~\cite{defossez2020real} on the Voicebank-DEMAND corpus~\cite{valentini2017noisy}. The mimic loss system uses a speech recognition model to provide a perceptual loss, achieving SotA performance on the COVL metric. Combining models for different tasks (as done here) is natural to implement in SpeechBrain. We also re-implemented the recently proposed MetricGAN+, which performs speech enhancement with an adversarially trained metric network  \cite{gan}.  Figure~\ref{fig:voicebank-evolution} shows the evolution of the PESQ performance on this corpus over the last few years. The SpeechBrain implementation of MetricGAN+ achieves the SotA PESQ performance when no extra data are used.

\begin{figure}
\CenterFloatBoxes
\begin{floatrow}
\capbtabbox{%
    \begin{tabular}{p{26mm}p{12mm}p{12mm}} \toprule
        \textbf{Technique} & \textbf{2-mix} & \textbf{3-mix}  \\ \midrule
   ConvTasnet  &  15.3 & 12.7 \\
   DualPath-RNN  &  18.8 & 14.7 \\
   SepFormer  &  20.4 & 17.6 \\
   SepFormer+DM  &  \textbf{22.3} & \textbf{19.5}\\ \bottomrule 
    \end{tabular}
}{%
  \caption{Scale-invariant signal-to-noise ratio improvement (SI-SNRi) in dB achieved with SpeechBrain on WSJ2mix and WSJ3mix.}
   \label{tab:separation}
}
\ffigbox{%
  \includegraphics[trim={0.6cm 0.6cm 1.6cm 1.4cm},clip,scale=0.43]{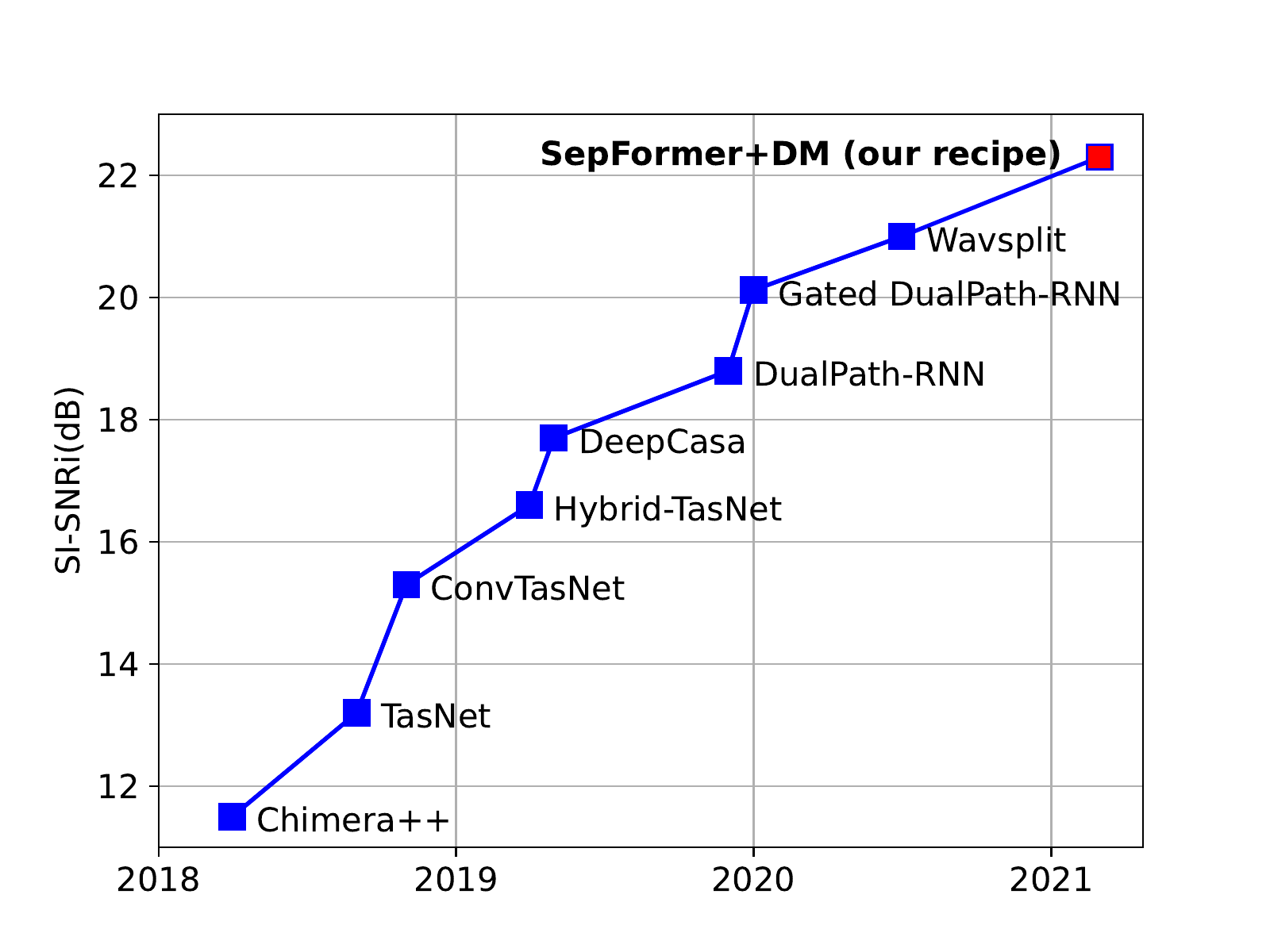}
}{%
  \caption{Evolution of the SotA performance (SI-SNRi) on the wsj2mix dataset. Source: \\\url{https://paperswithcode.com}.
  }
  \label{fig:wsj_soa}
}
\end{floatrow}
\end{figure}

SpeechBrain implements popular models for speech separation as well, namely ConvTasnet \cite{convtasnet} and Dual-path RNN \cite{luo2020dualpath}. Moreover, it supports the recently proposed SepFormer \cite{sepformer}, which uses a pipeline of two transformers within a dual-path framework. Table \ref{tab:separation} reports the results achieved on the standard WSJ0-2mix and WSJ0-3mix datasets \cite{deepclustering}, which contain mixtures composed of two or three overlapped speakers, respectively. The last row compares performance achieved with dynamic mixing, in which the training data are generated dynamically on-the-fly instead of using a frozen dataset. 
As shown in Figure~\ref{fig:wsj_soa}, SpeechBrain's SepFormer implementation achieves SotA on both datasets.


\section{Limitations and Future Work}

The current version of SpeechBrain supports many other tasks, including spoken language understanding, keyword spotting, multi-microphone signal processing, and language modeling. The toolkit also supports complex \cite{complex} and quaternion neural networks~\cite{qrnn}. Please refer to~\ref{sec:other_tasks} for further details. 
It does not currently support text-to-speech, which will be added shortly (pending pull-requests under review). 
In the future, we plan to support decoding with Finite State Transducers (FSTs) \cite{fst} and are considering to adopt the FST implementation of the ongoing k2 project \cite{k2} once stable. 
We plan to devote further effort to real-time speech processing, which was not the main focus of the first release. Finally, our goal is to add support for additional languages and further expand the set of recipes to open-source datasets not yet available in the toolkit (e.g., TED-LIUM \cite{tedlium}).

\section{Conclusion}
This paper described SpeechBrain, a novel, open-source, all-in-one speech processing toolkit. Our work illustrated the main design principles behind this toolkit and remarked on the design principles that led us to support multiple tasks without sacrificing simplicity, modularity, or flexibility.
Finally, we showed several use cases where the technology developed in SpeechBrain reaches SotA or competitive performance.
The main contribution to the scientific community is the development of a novel toolkit that can significantly accelerate future research in the fields of speech processing and deep learning. SpeechBrain is a coordinated effort towards making speech processing technology accessible, and are eager to see where its rapidly growing community of users takes the project in the future.


\begin{ack}
We would like to sincerely thank our generous sponsors: Samsung, Dolby, Nvidia, Nuance, ViaDialog. Special thanks to our institutional partners: Mila, LIA (Avignon University), CaMLSys (University of Cambridge), Sherbrooke University, and Bio-ASP (Academia Sinica). 
We also would like to acknowledge Breandan Considine, Olexa Bilaniuk, Frederic Osterrath, Mirko Bronzi, Anthony Larcher, Ralf Leibold, Salima Mdhaffar, Yannick Estève, Yu Tsao, Abdelmoumene Boumadane for helpful comments and discussions. We would like to express our gratitude to all the pre-release beta-testers and to the whole community that we are building around this project.
Thanks to Compute Canada for providing computational resources and support.
SpeechBrain was also granted access to the HPC resources of IDRIS under the allocation 2021-AD011012633 made by GENCI.
\end{ack}

\bibliography{mybib}

\begin{thebibliography}{113}
\providecommand{\natexlab}[1]{#1}
\providecommand{\url}[1]{\texttt{#1}}
\expandafter\ifx\csname urlstyle\endcsname\relax
  \providecommand{\doi}[1]{doi: #1}\else
  \providecommand{\doi}{doi: \begingroup \urlstyle{rm}\Url}\fi

\bibitem[Young et~al.(2002)Young, Evermann, Hain, Kershaw, Moore, Odell,
  Ollason, Povey, Valtchev, and Woodland]{Young02}
S.~Young, G.~Evermann, T.~Hain, D.~Kershaw, G.~Moore, J.~Odell, D.~Ollason,
  D.~Povey, V.~Valtchev, and P.~Woodland.
\newblock \emph{The {HTK} Book}.
\newblock Entropic Cambridge Research Laboratory, Cambridge, United Kingdom,
  2002.

\bibitem[Huggins-Daines et~al.(2006)Huggins-Daines, Kumar, Chan, Black,
  Ravishankar, and Rudnicky]{sphinx}
D.~Huggins-Daines, M.~Kumar, A.~Chan, A.~W. Black, M.~Ravishankar, and A.~I.
  Rudnicky.
\newblock Pocketsphinx: A free, real-time continuous speech recognition system
  for hand-held devices.
\newblock In \emph{in Proc. of ICASSP}, 2006.

\bibitem[Lee et~al.(2001)Lee, Kawahara, and Shikano]{julius}
A.~Lee, T.~Kawahara, and K.~Shikano.
\newblock Julius: An open source realtime large vocabulary recognition engine.
\newblock In \emph{Proc. of EUROSPEECH}, 2001.

\bibitem[Rybach et~al.(2011)Rybach, Hahn, Lehnen, Nolden, Sundermeyer,
  T{\"u}ske, Wiesler, Schl{\"u}ter, and Ney]{rwth}
D.~Rybach, S.~Hahn, P.~Lehnen, D.~Nolden, M.~Sundermeyer, Z.~T{\"u}ske,
  S.~Wiesler, R.~Schl{\"u}ter, and H.~Ney.
\newblock {RASR - The RWTH Aachen University Open Source Speech Recognition
  Toolkit}.
\newblock In \emph{Proc. of ASRU}, 2011.

\bibitem[Povey~et al.(2011)]{Povey}
D.~Povey~et al.
\newblock {The Kaldi Speech Recognition Toolkit}.
\newblock In \emph{Proc. of ASRU}, 2011.

\bibitem[Abadi et~al.(2016)Abadi, Barham, Chen, Chen, Davis, Dean, Devin,
  Ghemawat, Irving, Isard, Kudlur, Levenberg, Monga, Moore, Murray, Steiner,
  Tucker, Vasudevan, Warden, Wicke, Yu, and Zheng]{tensorflow}
M.~Abadi, P.~Barham, J.~Chen, Z.~Chen, A.~Davis, J.~Dean, M.~Devin,
  S.~Ghemawat, G.~Irving, M.~Isard, M.~Kudlur, J.~Levenberg, R.~Monga,
  S.~Moore, D.~G. Murray, B.~Steiner, P.~Tucker, V.~Vasudevan, P.~Warden,
  M.~Wicke, Y.~Yu, and X.~Zheng.
\newblock Tensorflow: A system for large-scale machine learning.
\newblock In \emph{Proc. of. USENIX Symposium on Operating Systems Design and
  Implementation}, 2016.

\bibitem[Paszke et~al.(2019)Paszke, Gross, Massa, Lerer, Bradbury, Chanan,
  Killeen, Lin, Gimelshein, Antiga, Desmaison, Kopf, Yang, DeVito, Raison,
  Tejani, Chilamkurthy, Steiner, Fang, Bai, and Chintala]{pytorch}
A.~Paszke, S.~Gross, F.~Massa, A.~Lerer, J.~Bradbury, G.~Chanan, T.~Killeen,
  Z.~Lin, N.~Gimelshein, L.~Antiga, A.~Desmaison, A.~Kopf, E.~Yang, Z.~DeVito,
  M.~Raison, A.~Tejani, S.~Chilamkurthy, B.~Steiner, L.~Fang, J.~Bai, and
  S.~Chintala.
\newblock Pytorch: An imperative style, high-performance deep learning library.
\newblock In \emph{Proc. of NeurIPS}, 2019.

\bibitem[Hannun et~al.(2014)Hannun, Case, Casper, Catanzaro, Diamos, Elsen,
  Prenger, Satheesh, Sengupta, Coates, and Ng]{deepspeech}
A.~Hannun, C.~Case, J.~Casper, B.~Catanzaro, G.~Diamos, E.~Elsen, R.~Prenger,
  S.~Satheesh, S.~Sengupta, A.~Coates, and A.~Y. Ng.
\newblock Deep speech: Scaling up end-to-end speech recognition, 2014.
\newblock arXiv:1412.5567.

\bibitem[Zeyer et~al.(2018)Zeyer, Alkhouli, and Ney]{returnn}
A.~Zeyer, T.~Alkhouli, and H.~Ney.
\newblock {RETURNN} as a generic flexible neural toolkit with application to
  translation and speech recognition.
\newblock In \emph{Proc. of ACML}, 2018.

\bibitem[Ravanelli et~al.(2019)Ravanelli, Parcollet, and Bengio]{pytorch_kaldi}
M.~Ravanelli, T.~Parcollet, and Y.~Bengio.
\newblock {The PyTorch-Kaldi Speech Recognition Toolkit}.
\newblock In \emph{Proc. of ICASSP}, 2019.

\bibitem[Wang et~al.(2019)Wang, Chen, Xu, Ding, Lv, Shao, Peng, Xie, Watanabe,
  and Khudanpur]{espresso}
Y.~Wang, T.~Chen, H.~Xu, S.~Ding, H.~Lv, Y.~Shao, N.~Peng, L.~Xie, S.~Watanabe,
  and S.~Khudanpur.
\newblock Espresso: A fast end-to-end neural speech recognition toolkit.
\newblock In \emph{Proc. of ASRU}, 2019.

\bibitem[Shen et~al.(2019)Shen, Nguyen, Wu, Chen, Chen, Jia, Kannan, Sainath,
  Cao, Chiu, et~al.]{lingvo}
J.~Shen, P.~Nguyen, Y.~Wu, Z.~Chen, M.~X. Chen, Y.~Jia, A.~Kannan, T.~Sainath,
  Y.~Cao, C.~Chiu, et~al.
\newblock Lingvo: A modular and scalable framework for sequence-to-sequence
  modeling.
\newblock 2019.
\newblock arXiv:1902.08295.

\bibitem[Ott et~al.(2019)Ott, Edunov, Baevski, Fan, Gross, Ng, Grangier, and
  Auli]{fairseq}
M.~Ott, S.~Edunov, A.~Baevski, A.~Fan, S.~Gross, N.~Ng, D.~Grangier, and
  M.~Auli.
\newblock Fairseq: A fast, extensible toolkit for sequence modeling, 2019.
\newblock arXiv:1904.01038.

\bibitem[Watanabe et~al.(2018)Watanabe, Hori, Karita, Hayashi, Nishitoba, Unno,
  Soplin, Heymann, Wiesner, Chen, Renduchintala, and Ochiai]{espnet}
S.~Watanabe, T.~Hori, S.~Karita, T.~Hayashi, J.~Nishitoba, Y.~Unno, N.~Soplin,
  J.~Heymann, M.~Wiesner, N.~Chen, A.~Renduchintala, and T.~Ochiai.
\newblock {ESPnet}: End-to-end speech processing toolkit.
\newblock In \emph{Proc. of Interspeech}, 2018.

\bibitem[Kuchaiev et~al.(2019)Kuchaiev, Li, Nguyen, Hrinchuk, Leary, Ginsburg,
  Kriman, Beliaev, Lavrukhin, Cook, Castonguay, Popova, Huang, and Cohen]{nemo}
O.~Kuchaiev, J.~Li, H.~Nguyen, O.~Hrinchuk, R.~Leary, B.~Ginsburg, S.~Kriman,
  S.~Beliaev, V.~Lavrukhin, J.~Cook, P.~Castonguay, M.~Popova, J.~Huang, and
  J.~M. Cohen.
\newblock {NeMo: a toolkit for building AI applications using Neural Modules},
  2019.
\newblock arXiv:1909.09577.

\bibitem[Pariente et~al.(2020)Pariente, Cornell, Cosentino, Sivasankaran,
  Tzinis, Heitkaemper, Olvera, Stöter, Hu, Martín-Doñas, Ditter, Frank,
  Deleforge, and Vincent]{asteroid}
M.~Pariente, S.~Cornell, J.~Cosentino, S.~Sivasankaran, E.~Tzinis,
  J.~Heitkaemper, M.~Olvera, F.~Stöter, M.~Hu, J.~M. Martín-Doñas,
  D.~Ditter, A.~Frank, A.~Deleforge, and E.~Vincent.
\newblock Asteroid: the {PyTorch}-based audio source separation toolkit for
  researchers.
\newblock In \emph{Proc. of Interspeech}, 2020.

\bibitem[Bredin et~al.(2020)Bredin, Yin, Coria, Gelly, Korshunov, Lavechin,
  Fustes, Titeux, Bouaziz, and Gill]{pyannote}
H.~Bredin, R.~Yin, J.~M. Coria, G.~Gelly, P.~Korshunov, M.~Lavechin, D.~Fustes,
  H.~Titeux, W.~Bouaziz, and M.~Gill.
\newblock {pyannote.audio: neural building blocks for speaker diarization}.
\newblock In \emph{Proc. of ICASSP}, 2020.

\bibitem[Larcher et~al.(2016)Larcher, Lee, and Meignier]{sidekit}
A.~Larcher, K.~A. Lee, and S.~Meignier.
\newblock An extensible speaker identification sidekit in python.
\newblock In \emph{Proc. of ICASSP}, 2016.

\bibitem[Yang et~al.(2021)Yang, Chi, Chuang, Lai, Lakhotia, Lin, Liu, Shi,
  Chang, Lin, Huang, Tseng, Lee, Liu, Huang, Dong, Li, Watanabe, Mohamed, and
  Lee]{superb}
S.~Yang, P.~Chi, Y.~Chuang, C.~Lai, K.~Lakhotia, Y.~Y. Lin, A.~T. Liu, J.~Shi,
  X.~Chang, G.~Lin, T.~Huang, W.~Tseng, K.~Lee, D.~Liu, Z.~Huang, S.~Dong,
  S.~Li, S.~Watanabe, A.~Mohamed, and H.~Lee.
\newblock Superb: Speech processing universal performance benchmark, 2021.
\newblock arXiv:2105.01051.

\bibitem[Wang and Wang(2016)]{jointraining}
Z.~Wang and D.~Wang.
\newblock A joint training framework for robust automatic speech recognition.
\newblock \emph{IEEE/ACM Transactions on Audio, Speech, and Language
  Processing}, 24\penalty0 (4):\penalty0 796--806, 2016.

\bibitem[Chen et~al.(2015)Chen, Watanabe, Erdogan, and Hershey]{Chen2015sep}
Z.~Chen, S.~Watanabe, H.~Erdogan, and J.R. Hershey.
\newblock Speech enhancement and recognition using multi-task learning of long
  short-term memory recurrent neural networks.
\newblock In \emph{Proc. of Interspeech}, 2015.

\bibitem[Gao et~al.(2015)Gao, Du, Dai, and Lee]{GaoDDL15}
T.~Gao, J.~Du, L.~Dai, and C.~Lee.
\newblock Joint training of front-end and back-end deep neural networks for
  robust speech recognition.
\newblock In \emph{Proc. of ICASSP}, 2015.

\bibitem[Ravanelli et~al.(2016)Ravanelli, Brakel, Omologo, and
  Bengio]{ravanelli_SLT}
M.~Ravanelli, P.~Brakel, M.~Omologo, and Y.~Bengio.
\newblock {Batch-normalized joint training for DNN-based distant speech
  recognition}.
\newblock In \emph{Proc. of SLT}, 2016.

\bibitem[Graves et~al.(2006)Graves, Fern{\'a}ndez, Gomez, and Schmidhuber]{CTC}
A.~Graves, S.~Fern{\'a}ndez, F.~Gomez, and J.~Schmidhuber.
\newblock Connectionist temporal classification: labelling unsegmented sequence
  data with recurrent neural networks.
\newblock In \emph{Proc. of ICML}, 2006.

\bibitem[Graves(2012)]{graves_transducer}
A.~Graves.
\newblock Sequence transduction with recurrent neural networks.
\newblock \emph{ICML --- Workshop on Representation Learning}, 2012.

\bibitem[Watanabe et~al.(2017)Watanabe, Hori, Kim, Hershey, and
  Hayashi]{hybrid_ctc_att}
S.~Watanabe, T.~Hori, S.~Kim, J.~R. Hershey, and T.~Hayashi.
\newblock Hybrid ctc/attention architecture for end-to-end speech recognition.
\newblock \emph{IEEE Journal of Selected Topics in Signal Processing},
  11\penalty0 (8):\penalty0 1240--1253, 2017.

\bibitem[Toshniwal et~al.(2018)Toshniwal, Kannan, Chiu, Wu, Sainath, and
  Livescu]{shallow_fusion}
S.~Toshniwal, A.~Kannan, C.~Chiu, Y.~Wu, T.~Sainath, and K.~Livescu.
\newblock A comparison of techniques for language model integration in
  encoder-decoder speech recognition.
\newblock 2018.

\bibitem[Panayotov et~al.(2015)Panayotov, Chen, Povey, and
  Khudanpur]{librispeech}
V.~Panayotov, G.~Chen, D.~Povey, and S.~Khudanpur.
\newblock {LibriSpeech: An ASR corpus based on public domain audio books}.
\newblock In \emph{Proc. of ICASSP}, 2015.

\bibitem[Ardila et~al.(2020)Ardila, Branson, Davis, Henretty, Kohler, Meyer,
  Morais, Saunders, Tyers, and Weber]{commonvoice:2020}
R.~Ardila, M.~Branson, K.~Davis, M.~Henretty, M.~Kohler, J.~Meyer, R.~Morais,
  L.~Saunders, F.~M. Tyers, and G.~Weber.
\newblock {Common Voice:} a massively-multilingual speech corpus.
\newblock In \emph{Proceedings of the 12th Conference on Language Resources and
  Evaluation (LREC 2020)}, 2020.

\bibitem[Bu et~al.(2017)Bu, Du, Na, Wu, and Zheng]{aishell_2017}
H.~Bu, J.~Du, X.~Na, B.~Wu, and H.~Zheng.
\newblock Aishell-1: An open-source mandarin speech corpus and a speech
  recognition baseline.
\newblock In \emph{Oriental COCOSDA}, 2017.

\bibitem[Garofolo et~al.(1993)Garofolo, Lamel, Fisher, Fiscus, Pallett, and
  Dahlgren]{timit}
J.~S. Garofolo, L.~F. Lamel, W.~M. Fisher, J.~G. Fiscus, D.~S. Pallett, and
  N.~L. Dahlgren.
\newblock {DARPA TIMIT Acoustic Phonetic Continuous Speech Corpus CDROM}, 1993.

\bibitem[Snyder et~al.(2018)Snyder, Garcia-Romero, Sell, Povey, and
  Khudanpur]{xvector}
D.~Snyder, D.~Garcia-Romero, G.~Sell, D.~Povey, and S.~Khudanpur.
\newblock X-vectors: Robust {DNN} embeddings for speaker recognition.
\newblock In \emph{Proc. of ICASSP}, 2018.

\bibitem[Desplanques et~al.(2020)Desplanques, Thienpondt, and Demuynck]{ecapa}
B.~Desplanques, J.~Thienpondt, and K.~Demuynck.
\newblock {ECAPA-TDNN:} emphasized channel attention, propagation and
  aggregation in {TDNN} based speaker verification.
\newblock In \emph{Proc. of Interspeech}, 2020.

\bibitem[Nagrani et~al.(2017)Nagrani, Chung, and Zisserman]{voxceleb}
A.~Nagrani, J.~S. Chung, and A.~Zisserman.
\newblock Voxceleb: a large-scale speaker identification dataset.
\newblock In \emph{Proc. of Interspech}, 2017.

\bibitem[Chung et~al.(2018)Chung, Nagrani, and Zisserman]{voxceleb2}
J.~S. Chung, A.~Nagrani, and A.~Zisserman.
\newblock Voxceleb2: Deep speaker recognition.
\newblock In \emph{Proc. of Interspeech}, 2018.

\bibitem[Luxburg(2007)]{spec_tutorial}
U.~Luxburg.
\newblock A tutorial on spectral clustering.
\newblock \emph{Statistics and Computing}, 17\penalty0 (4), 2007.

\bibitem[Dawalatabad et~al.(2021)Dawalatabad, Ravanelli, Grondin, Thienpondt,
  Desplanques, and Na]{ecapa_diarization}
N.~Dawalatabad, M.~Ravanelli, F.~Grondin, J.~Thienpondt, B.~Desplanques, and
  H.~Na.
\newblock {ECAPA-TDNN} embeddings for speaker diarization, 2021.
\newblock arXiv:2104.01466.

\bibitem[Carletta et~al.(2006)Carletta, Ashby, Bourban, Flynn, Guillemot, Hain,
  Kadlec, Karaiskos, Kraaij, Kronenthal, Lathoud, Lincoln, Lisowska, McCowan,
  Post, Reidsma, and Wellner]{ami-corpus}
J.~Carletta, S.~Ashby, S.~Bourban, M.~Flynn, M.~Guillemot, T.~Hain, J.~Kadlec,
  V.~Karaiskos, W.~Kraaij, M.~Kronenthal, G.~Lathoud, M.~Lincoln, A.~Lisowska,
  I.~McCowan, W.~Post, D.~Reidsma, and P.~Wellner.
\newblock The {AMI} meeting corpus: A pre-announcement.
\newblock In \emph{Proc. of the Second International Conference on Machine
  Learning for Multimodal Interaction}, 2006.

\bibitem[Fu et~al.(2021)Fu, Yu, Hsieh, Plantinga, Ravanelli, Lu, and
  Tsao]{fu2021metricgan+}
S.~Fu, C.~Yu, T.~Hsieh, P.~Plantinga, M.~Ravanelli, X.~Lu, and Y.~Tsao.
\newblock {MetricGAN+: An Improved Version of MetricGAN for Speech
  Enhancement}.
\newblock 2021.
\newblock arXiv:2104.03538.

\bibitem[Bagchi et~al.(2018)Bagchi, Plantinga, Stiff, and
  Fosler-Lussier]{bagchi2018spectral}
D.~Bagchi, P.~Plantinga, A.~Stiff, and E.~Fosler-Lussier.
\newblock Spectral feature mapping with mimic loss for robust speech
  recognition.
\newblock In \emph{Proc. of ICASSP}, 2018.

\bibitem[Veaux et~al.(2013)Veaux, Yamagishi, and King]{voicebank}
C.~Veaux, J.~Yamagishi, and S.~King.
\newblock The voice bank corpus: Design, collection and data analysis of a
  large regional accent speech database.
\newblock In \emph{International Conference Oriental COCOSDA held jointly with
  Conference on Asian Spoken Language Research and Evaluation
  (O-COCOSDA/CASLRE)}, 2013.

\bibitem[Reddy et~al.(2020)Reddy, Gopal, Cutler, Beyrami, Cheng, Dubey,
  Matusevych, Aichner, Aazami, Braun, Rana, Srinivasan, and
  Gehrke]{dns_challenge}
C.~Reddy, V.~Gopal, R.~Cutler, E.~Beyrami, R.~Cheng, H.~Dubey, S.~Matusevych,
  R.~Aichner, A.~Aazami, S.~Braun, P.~Rana, S.~Srinivasan, and J.~Gehrke.
\newblock The interspeech 2020 deep noise suppression challenge: Datasets,
  subjective testing framework, and challenge results.
\newblock In \emph{Proc. of Interspeech}, 2020.

\bibitem[Luo and Mesgarani(2019)]{convtasnet}
Y.~Luo and N.~Mesgarani.
\newblock Conv-tasnet: Surpassing ideal time-frequency magnitude masking for
  speech separation.
\newblock \emph{IEEE/ACM Transactions on Audio, Speech, and Language
  Processing}, 27\penalty0 (8):\penalty0 1256--1266, 2019.

\bibitem[Luo et~al.(2020)Luo, Chen, and Yoshioka]{luo2020dualpath}
Y.~Luo, Z.~Chen, and T.~Yoshioka.
\newblock {Dual-path RNN: efficient long sequence modeling for time-domain
  single-channel speech separation}, 2020.
\newblock arXiv:1910.06379.

\bibitem[Subakan et~al.(2021)Subakan, Ravanelli, Cornell, Bronzi, and
  Zhong]{sepformer}
C.~Subakan, M.~Ravanelli, S.~Cornell, M.~Bronzi, and J.~Zhong.
\newblock Attention is all you need in speech separation.
\newblock In \emph{Proc. of ICASSP}, 2021.

\bibitem[Hershey et~al.(2016)Hershey, Chen, {Le Roux}, and
  Watanabe]{deepclustering}
{J.} Hershey, Z.~Chen, J.~{Le Roux}, and S.~Watanabe.
\newblock Deep clustering: Discriminative embeddings for segmentation and
  separation.
\newblock In \emph{Proc. of ICASSP}, 2016.

\bibitem[Wichern et~al.(2019)Wichern, Antognini, Flynn, Zhu, McQuinn, Crow,
  Manilow, and Roux]{wham}
G.~Wichern, J.~Antognini, M.~Flynn, L.~Zhu, E.~McQuinn, D.~Crow, E.~Manilow,
  and J.~Le Roux.
\newblock {WHAM!: extending speech separation to noisy environments}.
\newblock In \emph{Proc. of Interspeech}, 2019.

\bibitem[Maciejewski et~al.(2020)Maciejewski, Wichern, McQuinn, and
  Roux]{whamr}
M.~Maciejewski, G.~Wichern, E.~McQuinn, and J.~Le Roux.
\newblock Whamr!: Noisy and reverberant single-channel speech separation.
\newblock In \emph{Proc. of ICASSP}, 2020.

\bibitem[Cosentino et~al.(2020)Cosentino, Pariente, Cornell, Deleforge, and
  Vincent]{librimix}
J.~Cosentino, M.~Pariente, S.~Cornell, A.~Deleforge, and E.~Vincent.
\newblock Librimix: An open-source dataset for generalizable speech separation,
  2020.
\newblock arXiv:2005.11262.

\bibitem[Lugosch et~al.(2021)Lugosch, Papreja, Ravanelli, Heba, and
  Parcollet]{timers-and-such}
L.~Lugosch, P.~Papreja, M.~Ravanelli, A.~Heba, and T.~Parcollet.
\newblock {Timers and Such: A} practical benchmark for spoken language
  understanding with numbers.
\newblock \emph{CoRR}, abs/2104.01604, 2021.

\bibitem[Haghani et~al.(2018)Haghani, Narayanan, Bacchiani, Chuang, Gaur,
  Moreno, Prabhavalkar, Qu, and Waters]{Haghani2018}
P.~Haghani, A.~Narayanan, M.~Bacchiani, G.~Chuang, N.~Gaur, P.~Moreno,
  R.~Prabhavalkar, Z.~Qu, and A.~Waters.
\newblock {From Audio to Semantics: Approaches to end-to-end spoken language
  understanding}.
\newblock \emph{IEEE Spoken Language Technology Workshop (SLT)}, 2018.

\bibitem[Serdyuk et~al.(2018)Serdyuk, Wang, Fuegen, Kumar, Liu, and
  Bengio]{Serdyuk2018}
D.~Serdyuk, Y.~Wang, C.~Fuegen, A.~Kumar, B.~Liu, and Y.~Bengio.
\newblock Towards end-to-end spoken language understanding.
\newblock In \emph{Proc. of ICASSP}, 2018.

\bibitem[Bastianelli et~al.(2020)Bastianelli, Vanzo, Swietojanski, and
  Rieser]{slurp}
E.~Bastianelli, A.~Vanzo, P.~Swietojanski, and V.~Rieser.
\newblock {SLURP: A Spoken Language Understanding Resource Package}.
\newblock In \emph{{Proc. of EMNLP}}, 2020.

\bibitem[Lugosch et~al.(2019)Lugosch, Ravanelli, Ignoto, Tomar, and
  Bengio]{fluent}
L.~Lugosch, M.~Ravanelli, P.~Ignoto, V.~Tomar, and Y.~Bengio.
\newblock Speech model pre-training for end-to-end spoken language
  understanding.
\newblock In \emph{Proc. of Interspeech}, 2019.

\bibitem[Habets et~al.(2009)Habets, Benesty, Cohen, Gannot, and
  Dmochowski]{habets2009new}
E.~Habets, J.~Benesty, I.~Cohen, S.~Gannot, and J.~Dmochowski.
\newblock New insights into the {MVDR} beamformer in room acoustics.
\newblock \emph{IEEE Transactions on Audio, Speech, and Language Processing},
  18\penalty0 (1):\penalty0 158--170, 2009.

\bibitem[Heymann et~al.(2016)Heymann, Drude, and
  Haeb-Umbach]{heymann2016neural}
J.~Heymann, L.~Drude, and R.~Haeb-Umbach.
\newblock Neural network based spectral mask estimation for acoustic
  beamforming.
\newblock In \emph{Proc. of ICASSP}, 2016.

\bibitem[Knapp and Carter(1976)]{KnappCarter}
C.~H. Knapp and G.~C. Carter.
\newblock The generalized correlation method for estimation of time delay.
\newblock \emph{IEEE Transactions on Acoustics, Speech, and Signal Processing},
  24\penalty0 (4):\penalty0 320--327, 1976.

\bibitem[Cobos et~al.(2010)Cobos, Marti, and Lopez]{cobos2010modified}
M.~Cobos, A.~Marti, and J.~Lopez.
\newblock A modified {SRP-PHAT} functional for robust real-time sound source
  localization with scalable spatial sampling.
\newblock \emph{IEEE Signal Processing Letters}, 18\penalty0 (1):\penalty0
  71--74, 2010.

\bibitem[Schmidt(1986)]{schmidt1986multiple}
R.~Schmidt.
\newblock Multiple emitter location and signal parameter estimation.
\newblock \emph{IEEE transactions on antennas and propagation}, 34\penalty0
  (3):\penalty0 276--280, 1986.

\bibitem[Gehring et~al.(2017)Gehring, Auli, Grangier, Yarats, and
  Dauphin]{convs2s}
J.~Gehring, M.~Auli, D.~Grangier, D.~Yarats, and Y.~Dauphin.
\newblock Convolutional sequence to sequence learning.
\newblock In \emph{Proc. of ICML}. PMLR, 2017.

\bibitem[Vaswani et~al.(2017)Vaswani, Shazeer, Parmar, Uszkoreit, Jones, Gomez,
  Kaiser, and Polosukhin]{transformers}
A.~Vaswani, N.~Shazeer, N.~Parmar, J.~Uszkoreit, L.~Jones, A.~N Gomez,
  {\L}.~Kaiser, and I.~Polosukhin.
\newblock Attention is all you need.
\newblock In \emph{Proc. of NeurIPS}, 2017.

\bibitem[Baevski et~al.(2020)Baevski, Zhou, Mohamed, and Auli]{wav2vect}
A.~Baevski, Y.~Zhou, A.~Mohamed, and M.~Auli.
\newblock wav2vec 2.0: A framework for self-supervised learning of speech
  representations.
\newblock In \emph{Proc. of NeurIPS}, 2020.

\bibitem[Li et~al.(2019)Li, Lavrukhin, Ginsburg, Leary, Kuchaiev, Cohen,
  Nguyen, and Gadde]{jasper}
J.~Li, V.~Lavrukhin, B.~Ginsburg, R.~Leary, O.~Kuchaiev, J.~M. Cohen,
  H.~Nguyen, and R.~Teja Gadde.
\newblock Jasper: An end-to-end convolutional neural acoustic model.
\newblock In \emph{Proc. of Interspeech}, 2019.

\bibitem[Kriman et~al.(2020)Kriman, Beliaev, Ginsburg, Huang, Kuchaiev,
  Lavrukhin, Leary, Li, and Zhang]{quartznet}
S.~Kriman, S.~Beliaev, B.~Ginsburg, J.~Huang, O.~Kuchaiev, V.~Lavrukhin,
  R.~Leary, J.~Li, and Y.~Zhang.
\newblock Quartznet: Deep automatic speech recognition with 1d time-channel
  separable convolutions.
\newblock In \emph{Proc. of ICASSP}, 2020.

\bibitem[Majumdar et~al.(2021)Majumdar, Balam, Hrinchuk, Lavrukhin, Noroozi,
  and Ginsburg]{citrinet}
S.~Majumdar, J.~Balam, O.~Hrinchuk, V.~Lavrukhin, V.~Noroozi, and B.~Ginsburg.
\newblock Citrinet: Closing the gap between non-autoregressive and
  autoregressive end-to-end models for automatic speech recognition, 2021.
\newblock arXiv:2104.01721.

\bibitem[Fairbanks et~al.(2006)Fairbanks, Garlan, and
  Scherlis]{fairbanks-2006-design-fragments}
G.~Fairbanks, D.~Garlan, and W.~Scherlis.
\newblock Design fragments make using frameworks easier.
\newblock \emph{SIGPLAN Not.}, 41\penalty0 (10), 2006.

\bibitem[Wolf et~al.(2020)Wolf, Debut, Sanh, Chaumond, Delangue, Moi, Cistac,
  Rault, Louf, Funtowicz, Davison, Shleifer, von Platen, Ma, Jernite, Plu, Xu,
  Le~Scao, Gugger, Drame, Lhoest, and Rush]{hugging}
T.~Wolf, L.~Debut, V.~Sanh, J.~Chaumond, C.~Delangue, A.~Moi, P.~Cistac,
  T.~Rault, R.~Louf, M.~Funtowicz, J.~Davison, S.~Shleifer, P.~von Platen,
  C.~Ma, Y.~Jernite, J.~Plu, C.~Xu, T.~Le~Scao, S.~Gugger, M.~Drame, Q.~Lhoest,
  and A.~Rush.
\newblock Transformers: State-of-the-art natural language processing.
\newblock In \emph{Proc of EMNLP}, 2020.

\bibitem[Variani et~al.(2017)Variani, Bagby, McDermott, and
  Bacchiani]{batching}
E.~Variani, T.~Bagby, E.~McDermott, and M.~Bacchiani.
\newblock End-to-end training of acoustic models for large vocabulary
  continuous speech recognition with tensorflow.
\newblock In \emph{Proc. of Interspeech}, 2017.

\bibitem[Morishita et~al.(2017)Morishita, Oda, Neubig, Yoshino, Sudoh, and
  Nakamura]{morishita-etal-2017-empirical}
M.~Morishita, Y.~Oda, G.~Neubig, K.~Yoshino, K.~Sudoh, and S.~Nakamura.
\newblock An empirical study of mini-batch creation strategies for neural
  machine translation.
\newblock In \emph{Proc. of the WNMT}, 2017.

\bibitem[Pedregosa et~al.(2011)Pedregosa, Varoquaux, Gramfort, Michel, Thirion,
  Grisel, Blondel, Prettenhofer, Weiss, Dubourg, et~al.]{pedregosa2011scikit}
F.~Pedregosa, G.~Varoquaux, A.~Gramfort, V.~Michel, B.~Thirion, O.~Grisel,
  M.~Blondel, P.~Prettenhofer, R.~Weiss, V.~Dubourg, et~al.
\newblock Scikit-learn: Machine learning in python.
\newblock \emph{the Journal of machine Learning research}, 12:\penalty0
  2825--2830, 2011.

\bibitem[Virtanen et~al.(2020)Virtanen, Gommers, Oliphant, Haberland, Reddy,
  Cournapeau, Burovski, Peterson, Weckesser, Bright, et~al.]{virtanen2020scipy}
P.~Virtanen, R.~Gommers, T.~E Oliphant, M.~Haberland, T.~Reddy, David
  Cournapeau, Evgeni Burovski, Pearu Peterson, Warren Weckesser, Jonathan
  Bright, et~al.
\newblock {SciPy} 1.0: fundamental algorithms for scientific computing in
  {Python}.
\newblock \emph{Nature methods}, 17\penalty0 (3):\penalty0 261--272, 2020.

\bibitem[Gulli and Pal(2017)]{gulli2017deep}
A.~Gulli and S.~Pal.
\newblock \emph{Deep learning with {Keras}}.
\newblock Packt Publishing Ltd, 2017.

\bibitem[Howard and Gugger(2020)]{howard2020fastai}
J.~Howard and S.~Gugger.
\newblock Fastai: A layered api for deep learning.
\newblock \emph{Information}, 11\penalty0 (2):\penalty0 108, 2020.

\bibitem[Falcon et~al.(2019)]{falcon2019pytorch}
W.~Falcon et~al.
\newblock Pytorch {Lightning}.
\newblock \emph{GitHub}, 2019.
\newblock \url{https://github.com/PyTorchLightning/pytorch-lightning}.

\bibitem[Goodfellow et~al.(2014)Goodfellow, Pouget-Abadie, Mirza, Xu,
  Warde-Farley, Ozair, Courville, and Bengio]{gan}
I.~Goodfellow, J.~Pouget-Abadie, M.~Mirza, B.~Xu, D.~Warde-Farley, S.~Ozair,
  A.~Courville, and Y.~Bengio.
\newblock Generative adversarial nets.
\newblock \emph{Proc. of NIPS}, 2014.

\bibitem[Park et~al.(2019)Park, Chan, Zhang, Chiu, Zoph, Cubuk, and
  Le]{Park2019}
D.~Park, W.~Chan, Y.~Zhang, C.~Chiu, B.~Zoph, E.~Cubuk, and Q.~Le.
\newblock {SpecAugment: A Simple Data Augmentation Method for Automatic Speech
  Recognition}.
\newblock In \emph{Proc. of Interspeech}, 2019.

\bibitem[Graves and Jaitly(2014)]{graves_ctc}
A.~Graves and N.~Jaitly.
\newblock End-to-end speech recognition with recurrent neural networks.
\newblock In \emph{Proc. of ICML}, 2014.

\bibitem[Hochreiter and Schmidhuber(1997)]{lstm}
S.~Hochreiter and J.~Schmidhuber.
\newblock Long short-term memory.
\newblock \emph{Neural Computation}, 9\penalty0 (8):\penalty0 1735--1780, 1997.

\bibitem[Chung et~al.(2014)Chung, G{\"{u}}l{\c c}ehre, Cho, and Bengio]{gru2}
J.~Chung, {\c C}.~G{\"{u}}l{\c c}ehre, K.~Cho, and Y.~Bengio.
\newblock Empirical evaluation of gated recurrent neural networks on sequence
  modeling.
\newblock In \emph{Proc. of NIPS}, 2014.

\bibitem[Ravanelli et~al.(2018)Ravanelli, Brakel, Omologo, and Bengio]{li_gru}
M.~Ravanelli, P.~Brakel, M.~Omologo, and Y.~Bengio.
\newblock Light gated recurrent units for speech recognition.
\newblock \emph{IEEE Transactions on Emerging Topics in Computational
  Intelligence}, 2\penalty0 (2):\penalty0 92--102, 2018.

\bibitem[Kudo and Richardson(2018)]{sentencepiece}
T.~Kudo and J.~Richardson.
\newblock {S}entence{P}iece: A simple and language independent subword
  tokenizer and detokenizer for neural text processing.
\newblock In \emph{Proc. of EMNLP}, 2018.

\bibitem[Kahn et~al.(2019)Kahn, Rivi{\`{e}}re, Zheng, Kharitonov, Xu,
  Mazar{\'{e}}, Karadayi, Liptchinsky, Collobert, Fuegen, Likhomanenko,
  Synnaeve, Joulin, Mohamed, and Dupoux]{librilight}
J.~Kahn, M.~Rivi{\`{e}}re, W.~Zheng, E.~Kharitonov, Q.~Xu, P.~Mazar{\'{e}},
  J.~Karadayi, V.~Liptchinsky, R.~Collobert, C.~Fuegen, T.~Likhomanenko,
  G.~Synnaeve, A.~Joulin, A.~Mohamed, and E.~Dupoux.
\newblock Libri-light: {A} benchmark for {ASR} with limited or no supervision.
\newblock \emph{CoRR}, abs/1912.07875, 2019.

\bibitem[Han et~al.(2020)Han, Zhang, Zhang, Yu, Chiu, Qin, Gulati, Pang, and
  Wu]{contextnet}
W.~Han, Z.~Zhang, Y.~Zhang, J.~Yu, C.~Chiu, J.~Qin, A.~Gulati, R.~Pang, and
  Y.~Wu.
\newblock Contextnet: Improving convolutional neural networks for automatic
  speech recognition with global context.
\newblock 2020.
\newblock arXiv:2005.03191.

\bibitem[Karita et~al.(2019)Karita, Chen, Hayashi, Hori, Inaguma, Jiang,
  Someki, Soplin, Yamamoto, Wang, Watanabe, Yoshimura, and
  Zhang]{transformers_asr}
S.~Karita, N.~Chen, T.~Hayashi, T.~Hori, H.~Inaguma, Z.~Jiang, M.~Someki,
  N.~Soplin, R.~Yamamoto, X.~Wang, S.~Watanabe, T.~Yoshimura, and W.~Zhang.
\newblock A comparative study on transformer vs rnn in speech applications.
\newblock In \emph{Proc. of ASRU}, 2019.

\bibitem[Szyma{\'n}ski et~al.(2020)Szyma{\'n}ski, {\.Z}elasko, Morzy, Szymczak,
  {\.Z}y{\l}a-Hoppe, Banaszczak, Augustyniak, Mizgajski, and
  Carmiel]{szymanski-etal-2020-wer}
P.~Szyma{\'n}ski, P.~{\.Z}elasko, M.~Morzy, A.~Szymczak, M.~{\.Z}y{\l}a-Hoppe,
  J.~Banaszczak, L.~Augustyniak, J.~Mizgajski, and Y.~Carmiel.
\newblock {WER} we are and {WER} we think we are.
\newblock In \emph{Proc. of EMNLP}, 2020.

\bibitem[Likhomanenko et~al.(2020)Likhomanenko, Xu, Pratap, Tomasello, Kahn,
  Avidov, Collobert, and Synnaeve]{rethinking}
T.~Likhomanenko, Q.~Xu, V.~Pratap, P.~Tomasello, J.~Kahn, G.~Avidov,
  R.~Collobert, and G.~Synnaeve.
\newblock Rethinking evaluation in {ASR:} are our models robust enough?
\newblock \emph{CoRR}, abs/2010.11745, 2020.

\bibitem[Zeinali et~al.(2019)Zeinali, Wang, Silnova, Mat\v{e}jka, and
  Plchot]{FITPUB12224}
H.~Zeinali, S.~Wang, A.~Silnova, P.~Mat\v{e}jka, and O.~Plchot.
\newblock But system description to voxceleb speaker recognition challenge
  2019.
\newblock In \emph{Proc. of The VoxCeleb Challange Workshop}, 2019.

\bibitem[Pal et~al.(2020)Pal, Kumar, Peri, Park, Kim, Lord, Bishop, and
  Narayanan]{pal21-meta}
M.~Pal, M.~Kumar, R.~Peri, T.~Park, S.~Kim, C.~Lord, S.~Bishop, and
  S.~Narayanan.
\newblock Meta-learning with latent space clustering in generative adversarial
  network for speaker diarization, 2020.
\newblock arXiv:2007.09635.

\bibitem[Landini et~al.(2020)Landini, Profant, Diez, and
  Burget]{landini2020VBX}
F.~Landini, J.~Profant, M.~Diez, and L.~Burget.
\newblock Bayesian {HMM} clustering of x-vector sequences ({VBx}) in speaker
  diarization: theory, implementation and analysis on standard tasks, 2020.
\newblock arXiv:2012.14952.

\bibitem[Wang et~al.(2021)Wang, Rivi{\`e}re, Lee, Wu, Talnikar, Haziza,
  Williamson, Pino, and Dupoux]{wang2021voxpopuli}
C.~Wang, M.~Rivi{\`e}re, A.~Lee, A.~Wu, C.~Talnikar, D.~Haziza, M.~Williamson,
  J.~Pino, and E.~Dupoux.
\newblock Voxpopuli: A large-scale multilingual speech corpus for
  representation learning, semi-supervised learning and interpretation.
\newblock \emph{arXiv:2101.00390}, 2021.

\bibitem[Lang and Hinton(1988)]{Lang+Hinton88}
K.~J. Lang and G.~E. Hinton.
\newblock The development of the time-delay neural network architecture for
  speech recognition.
\newblock Technical Report CMU-CS-88-152, Carnegie-Mellon University, 1988.

\bibitem[Waibel et~al.(1989)Waibel, Hanazawa, Hinton, Shikano, and
  Lang]{waibel}
A.~Waibel, T.~Hanazawa, G.~Hinton, K.~Shikano, and K.~Lang.
\newblock Phoneme recognition using time-delay neural networks.
\newblock \emph{IEEE Transactions on Acoustics, Speech, and Signal Processing},
  37:\penalty0 328--339, 1989.

\bibitem[Kenny et~al.(2013)Kenny, Stafylakis, Ouellet, Alam, and
  Dumouchel]{Kenny-plda}
P.~Kenny, T.~Stafylakis, P.~Ouellet, Md.~J. Alam, and P.~Dumouchel.
\newblock {PLDA} for speaker verification with utterances of arbitrary
  duration.
\newblock In \emph{Proc. of ICASSP}, 2013.

\bibitem[Garcia-Romero and Espy-Wilson(2011)]{GarciaRomero2011AnalysisOI}
D.~Garcia-Romero and C.~Espy-Wilson.
\newblock Analysis of i-vector length normalization in speaker recognition
  systems.
\newblock In \emph{Proc. of Interspeech}, 2011.

\bibitem[D{\'e}fossez et~al.(2020)D{\'e}fossez, Synnaeve, and
  Adi]{defossez2020real}
A.~D{\'e}fossez, G.~Synnaeve, and Y.~Adi.
\newblock Real time speech enhancement in the waveform domain.
\newblock \emph{Proc. of Interspeech}, 2020.

\bibitem[Valentini-Botinhao(2017)]{valentini2017noisy}
C.~Valentini-Botinhao.
\newblock Noisy speech database for training speech enhancement algorithms and
  {TTS} models.
\newblock \emph{Edinburgh DataShare}, 2017.

\bibitem[Trabelsi et~al.(2018)Trabelsi, Bilaniuk, Zhang, Serdyuk, Subramanian,
  Santos, Mehri, Rostamzadeh, Bengio, and Pal]{complex}
C.~Trabelsi, O.~Bilaniuk, Y.~Zhang, D.~Serdyuk, S.~Subramanian, J.~Santos,
  S.~Mehri, N.~Rostamzadeh, Y.~Bengio, and C.~Pal.
\newblock Deep complex networks.
\newblock In \emph{Proc. of ICLR}, 2018.

\bibitem[Parcollet et~al.(2019)Parcollet, Ravanelli, Morchid, Linar{\`{e}}s,
  Trabelsi, Mori, and Bengio]{qrnn}
T.~Parcollet, M.~Ravanelli, M.~Morchid, G.~Linar{\`{e}}s, C.~Trabelsi, R.~De
  Mori, and Y.~Bengio.
\newblock Quaternion recurrent neural networks.
\newblock In \emph{Proc. of ICLR}, 2019.

\bibitem[Mohri et~al.(2002)Mohri, Pereira, and Riley]{fst}
M.~Mohri, F.~Pereira, and M.~Riley.
\newblock Weighted finite-state transducers in speech recognition.
\newblock \emph{Computer Speech and Language}, 16\penalty0 (1):\penalty0
  69–88, 2002.

\bibitem[Povey et~al.(2020)]{k2}
D.~Povey et~al.
\newblock k2.
\newblock \url{https://github.com/k2-fsa/k2}, 2020.

\bibitem[Hernandez et~al.(2018)Hernandez, Nguyen, Ghannay, Tomashenko, and
  Est{\`{e}}ve]{tedlium}
F.~Hernandez, V.~Nguyen, S.~Ghannay, N.~Tomashenko, and Y.~Est{\`{e}}ve.
\newblock {TED-LIUM} 3: Twice as much data and corpus repartition for
  experiments on speaker adaptation.
\newblock In \emph{Proc. of SPECOM}, 2018.

\bibitem[Li et~al.(2021)Li, Shi, Zhang, Subramanian, Chang, Kamo, Hira,
  Hayashi, B{\"{o}}ddeker, Chen, and Watanabe]{espnet_se}
C.~Li, J.~Shi, W.~Zhang, A.~Subramanian, X.~Chang, N.~Kamo, M.~Hira,
  T.~Hayashi, C.~B{\"{o}}ddeker, Z.~Chen, and Shinji Watanabe.
\newblock Espnet-se: End-to-end speech enhancement and separation toolkit
  designed for {ASR} integration.
\newblock In \emph{Proc. of the IEEE Spoken Language Technology Workshop},
  2021.

\bibitem[Erdogan et~al.(2016)Erdogan, Hershey, Watanabe, Mandel, and
  Le~Roux]{erdogan2016improved}
H.~Erdogan, J.~Hershey, S.~Watanabe, M.~Mandel, and J.~Le~Roux.
\newblock Improved {MVDR} beamforming using single-channel mask prediction
  networks.
\newblock In \emph{Proc. of Interspeech}, 2016.

\bibitem[Grondin et~al.(2020)Grondin, Lauzon, Vincent, and
  Michaud]{grondin2020gev}
F.~Grondin, J.~Lauzon, J.~Vincent, and F.~Michaud.
\newblock {GEV} beamforming supported by {DOA}-based masks generated on pairs
  of microphones.
\newblock In \emph{Proc. of Interspeech}, 2020.

\bibitem[Grondin and Michaud(2019)]{grondin2019lightweight}
F.~Grondin and F.~Michaud.
\newblock Lightweight and optimized sound source localization and tracking
  methods for open and closed microphone array configurations.
\newblock \emph{Robotics and Autonomous Systems}, 2019.

\bibitem[Omologo and Svaizer(1994)]{gcf}
M.~Omologo and P.~Svaizer.
\newblock Acoustic event localization using a crosspower-spectrum phase based
  technique.
\newblock In \emph{Proc. of ICASSP}, 1994.

\bibitem[Du et~al.(2014)Du, Wang, Gao, Xu, Dai, and Lee]{du2014robust}
J.~Du, Q.~Wang, T.~Gao, Y.~Xu, L.~Dai, and C.~Lee.
\newblock Robust speech recognition with speech enhanced deep neural networks.
\newblock In \emph{Proc. of Interspeech}, 2014.

\bibitem[Bahdanau et~al.(2015)Bahdanau, Cho, and Bengio]{bahdanau2014neural}
D.~Bahdanau, K.~Cho, and Y.~Bengio.
\newblock Neural machine translation by jointly learning to align and
  translate.
\newblock \emph{Proc. of ICLR}, 2015.

\bibitem[Seo et~al.(2021)Seo, Kwak, and Lee]{seo2021integration}
S.~Seo, D.~Kwak, and B.~Lee.
\newblock Integration of pre-trained networks with continuous token interface
  for end-to-end spoken language understanding.
\newblock \emph{arXiv:2104.07253}, 2021.

\bibitem[Thiemann et~al.(2013)Thiemann, Ito, and Vincent]{demand}
J.~Thiemann, N.~Ito, and E.~Vincent.
\newblock {The Diverse Environments Multi-channel Acoustic Noise Database
  (DEMAND): A database of multichannel environmental noise recordings}.
\newblock In \emph{{21st International Congress on Acoustics}}, 2013.

\bibitem[Rix et~al.(2001)Rix, Beerends, Hollier, and Hekstra]{pesq}
A.W. Rix, J.G. Beerends, M.P. Hollier, and A.P. Hekstra.
\newblock Perceptual evaluation of speech quality (pesq)-a new method for
  speech quality assessment of telephone networks and codecs.
\newblock In \emph{Proc. of ICASSP}, 2001.

\bibitem[Hu and Loizou(2007)]{hu2007evaluation}
Y.~Hu and P.~Loizou.
\newblock Evaluation of objective quality measures for speech enhancement.
\newblock \emph{IEEE Transactions on Audio, Speech, and Language Processing},
  16\penalty0 (1):\penalty0 229--238, 2007.

\bibitem[Vincent et~al.(2006)Vincent, Gribonval, and Fevotte]{bsseval}
E.~Vincent, R.~Gribonval, and C.~Fevotte.
\newblock Performance measurement in blind audio source separation.
\newblock \emph{IEEE Transactions on Audio, Speech, and Language Processing},
  14\penalty0 (4):\penalty0 1462--1469, 2006.

\end{thebibliography}

\newpage
\appendix

\section{Appendix}
\subsection{Statement on social impact}
Speech technologies can support humans in a variety of positive ways (e.g., helping hearing-impaired individuals, detecting speech pathologies, helping people learning new languages, allowing people with physical disabilities to control their home appliances, etc.). They can make our life safer (e.g., in-car speech recognition) or just more comfortable (e.g., with voice assistants, etc.). The growing demand for speech technology observed in the last few years confirms the importance of this technology in everyday lives.  
However, non-ethical misuses of these technologies are possible as well. Most of them are related to well-known privacy concerns, which can be mitigated with more rigid regulations such as the General Data Protection Regulation\footnote{\url{https://gdpr-info.eu}} (GDPR) adopted in Europe.

As with all other open-source toolkits, we cannot have full control over the actual use of the developed technologies.  However, we strongly encourage ethical use of our toolkit,  and we ask all SpeechBrain users to fully respect the \textit{Montreal Declaration for a Responsible Development of Artificial Intelligence}\footnote{\url{https://www.montrealdeclaration-responsibleai.com/}}. 
Moreover, we think that having open-source technology available to everyone is better than leaving it in the hands of a few players only. This can potentially mitigate the negative consequences of this ongoing societal change towards an AI-aided society.

\subsection{Performance comparison with other toolkits}
\label{app:comp_toolkits}
Comparing the performance across speech processing toolkits is often problematic for several reasons and can be deceptive if not framed in a much larger context. 
First, each toolkit focuses more on some tasks or models and provides recipes only for specific datasets. 
Secondly, there are intrinsic differences in how these toolkits implement recipes for the same task on the same dataset. For example, Kaldi relies only on hybrid speech recognition, while others such as SpeechBrain, ESPNet, and NeMo do not currently support hybrid speech recognition but instead provide more modern E2E speech recognition models. 
Thirdly, even across recipes concerning the same model on the same dataset, some differences arise due to different feature extraction, data loading pipelines, batching mechanisms, and other implementation details. Finally, most of the aforementioned toolkits are active projects, and the performance of a given task might change over time. 
We think that the comparison proposed in the section can only be used to probe whether a toolkit can provide reasonable performance compared to other open-source implementations.
Table \ref{tab:comparison_other_toolkits} compares the best results reported in the official repository of each toolkit on tasks and datasets we have found in common (as of May 2021).

\begin{table}[!htbp]
    \centering
    \setlength{\tabcolsep}{7pt} 
    \renewcommand{\arraystretch}{1} 
    \footnotesize
    \begin{tabular}{lllcccc}
         {\bf Task type} & {\bf Metric} & {\bf Dataset} & { \bf SpeechBrain} &  { \bf ESPNet} & {\bf NeMo} & {\bf Kaldi} \\
         \hline
                Speech Rec. & WER(\%) $\downarrow$ & Common Voice Fr & \bf{13.34}* & 13.9\footnote{\url{https://github.com/espnet/espnet/blob/master/egs2/commonvoice/asr1/README.md}} $\dagger$ & 14.01\footnote{\url{https://ngc.nvidia.com/catalog/models/nvidia:nemo:stt_fr_quartznet15x5}} & n.a \\
                Speech Rec. & WER(\%) $\downarrow$ & Common Voice It & \bf{9.86}* & 16.1\footnote{\url{https://github.com/espnet/espnet/blob/master/egs2/commonvoice/asr1/README.md}} $\dagger$ & 15.22\footnote{\url{https://ngc.nvidia.com/catalog/models/nvidia:nemo:stt_it_quartznet15x5}} & n.a \\
                Speech Rec. & WER(\%) $\downarrow$ & LibriSpeech \textbf{\lstinline{test-clean}} & 2.46 & 2.1 \footnote{\url{https://github.com/espnet/espnet/tree/master/egs2/librispeech/asr1}}  & \textbf{2.00}\footnote{Result taken from \cite{citrinet}.} & 4.17\footnote{\url{https://github.com/kaldi-asr/kaldi/blob/master/egs/librispeech/s5/RESULTS}} \\ 
                 Speech Rec.   & CER(\%) $\downarrow$ &  AISHELL-1 & 5.58 & \bf{4.7}\footnote{\url{https://github.com/espnet/espnet/tree/master/egs2/aishell/asr1}}$\dagger$ & 5.55\footnote{Results taken from \cite{citrinet}. It uses extra data from Multilingual LibriSpeech.} & 7.43\footnote{\url{https://github.com/kaldi-asr/kaldi/blob/master/egs/aishell/s5/RESULTS}} \\ 
                    Speech Rec. & PER(\%) $\downarrow$ &  TIMIT & \bf{8.04}* & 19.5\footnote{\url{https://github.com/espnet/espnet/blob/master/egs2/timit/asr1/README.md}} & n.a. & 18.4\footnote{\url{https://github.com/kaldi-asr/kaldi/blob/master/egs/timit/s5/RESULTS}} \\
                    Speaker Ver.  & EER(\%) $\downarrow$ & Voxceleb1+2 & \textbf{0.69} & n.a & 2.05\footnote{\url{https://ngc.nvidia.com/catalog/models/nvidia:nemo:speakerverification_speakernet/}} & 3.10 \footnote{\url{https://kaldi-asr.org/models/m7}} \\ 
                  Speech Sep.  & SNRi(dB) $\uparrow$ & WSJ2-mix & \bf{22.3} & 17.9\footnote{Results taken from \cite{espnet_se}.} & n.a. & n.a. \\ 
        \hline
    \end{tabular}
    \caption{Performance comparison across speech toolkits on common tasks. For each toolkit, dataset, and task we report the best performance on the test set (as of May 2021). The arrow $\downarrow$ indicates the lower the better, while $\uparrow$ indicates the higher the better.}
    \label{tab:comparison_other_toolkits}
    \vspace{1ex}
{\raggedright \footnotesize{*uses self-supervised pre-training with wav2vec 2.0.} \par} 
{\raggedright \footnotesize{$\dagger$ ESPNet uses transformer language models  (SpeechBrain does not for these tasks)}.\par}    
\end{table}

We can see that SpeechBrain achieves competitive performance with other pre-existing toolkits across different tasks and datasets. 
It is worth mentioning that all the toolkits considered here can support all the tasks and datasets reported in Table \ref{tab:comparison_other_toolkits}.  Each toolkit can potentially fill the performance gap with the best-performing one just by implementing a better model with properly fine-tuned hyperparameters for the specific task. We thus think that the actual value of a toolkit mainly lies in its usability and flexibility, which are the main principles that guided the design of SpeechBrain.

\subsection{Additional tasks}
In the following, we describe some of the supported applications not discussed in the main paper. 
\label{sec:other_tasks}
\subsubsection{Multi-microphone signal processing}

Multi-microphone signal processing techniques are useful in different ways within a speech processing pipeline.
The information captured by different microphones can be used to estimate the direction of arrival (DOA) of a sound source. We can then use beamforming to enhance the target source.
SpeechBrain performs multi-channel processing in the frequency domain.
For both DOA estimation and beamforming, it is assumed that the spatial covariance matrices (SCMs) are computed for each frequency bin $k$ using the Short-Time Fourier Transform (STFT).
We denote the SCMs for the target speech, the interfering noise and the resulting mixture as $\mathbf{R}_{SS}[k] \in \mathbb{C}^{M \times M}$, $\mathbf{R}_{NN}[k] \in \mathbb{C}^{M \times M}$ and $\mathbf{R}_{XX}[k] \in \mathbb{C}^{M \times M}$, respectively, where $M$ stands for the number of microphones.
The SCMs for speech and noise can be obtained using time-frequency masks \cite{heymann2016neural,erdogan2016improved,grondin2020gev}.

The DOA of the sound can be computed using the Generalized Cross-Correlation Phase Transform (GCC-PHAT) \cite{KnappCarter},  the Steered-Response Power with Phase Transform (SRP-PHAT) 
\cite{cobos2010modified}, or the the Multiple Signal Classification (MUSIC) algorithm. All of these techniques are implemented in SpeechBrain using GPU-friendly functions.
The GCC-PHAT computes the DOA on a pair of microphones and returns the time difference of arrival (TDOA), which can be mapped to a DOA on an arc from $0^{\circ}$ to $180^{\circ}$.
SRP-PHAT scans each potential DOA on a virtual unit sphere around the array and computes the corresponding power \cite{grondin2019lightweight}. 
For each DOA (denoted by the unit vector $\mathbf{u}$), there is a steering vector $\mathbf{A}(k,\mathbf{u}) \in \mathbb{C}^{M \times 1}$ in the direction of $\mathbf{u}$:

\begin{equation}
E(\mathbf{u}) = \sum_{p=1}^{M}{\sum_{q=p+1}^{M}{\sum_{k}{\frac{R_{p,q}[k]}{|R_{p,q}[k]|}A_p(k,\mathbf{u})A_q(k,\mathbf{u})^*}}}
\end{equation}
where $R_{p,q}[k]$ stands for the element at the $p$-th row and $q$-th column in the SCM, and $A_p(k,\mathbf{u})$ and $A_q(k,\mathbf{u})$ stands for the $p$-th and $q$-th elements of the steering vector.
The DOA $\mathbf{u}$ with the maximum power $E(\mathbf{u})$ is selected as the DOA of sound. It is worth mentioning that SRP-PHAT \cite{cobos2010modified} is conceptually the same as another popular localization technique called Global Coherence Field (GCF) \cite{gcf}, which projects the DOA information into 2D or 3D plans. That will be possibly the object of future implementation in SpeechBrain.

It is also possible to estimate the DOA using the Multiple Signal Classification (MUSIC) algorithms \cite{schmidt1986multiple}.
MUSIC scans each potential direction of arrival on a virtual unit sphere around the array and computes the corresponding power.
The matrix $\mathbf{U}(k) \in \mathbb{C}^{M \times S}$ contains the $S$ eigenvectors that correspond to the $S$ smallest eigenvalues obtained while performing eigenvalue decomposition on the SCM. The power corresponds to: 
\begin{equation}
E(\mathbf{u}) = \sum_{k}\frac{\mathbf{A}(k,\mathbf{u})^H \mathbf{A}(k,\mathbf{u})}{\sqrt{\mathbf{A}(k,\mathbf{u})^H \mathbf{U}(k)\mathbf{U}(k)^H\mathbf{A}(k,\mathbf{u})}}
\end{equation}
where $\{\dots\}^H$ stands for the Hermitian operator, and the DOA corresponds to the unit vector $\mathbf{u}$ associated with the maximum value of $E(\mathbf{u})$.

Speech can be enhanced with beamforming methods.
The most straightforward approach consists of using a delay-and-sum beamformer to produce constructive interference in the DOA of the target sound source.
Beamforming generates frequency-wise coefficients $\mathbf{W}(k) \in \mathbb{C}^{M \times 1}$ that multiply the STFT of each microphone ($\mathbf{X}(t,k) \in \mathbb{C}^{M \times T}$) and adds the products to produce the enhanced speech STFT ($Y(t,k) \in \mathbb{C}^{1 \times T}$):

\begin{equation}
    Y(t,k) = \mathbf{W}^H(k) \mathbf{X}(t,k)
\end{equation}

With delay-and-sum, the coefficients are obtained using the steering vector as follows:

\begin{equation}
\mathbf{W}(k) = \frac{1}{M}\mathbf{A}(k)    
\end{equation}

Alternatively, the Minimum Variance Distortionless Response (MVDR) beamformer \cite{habets2009new} exploits the DOA but also the SCMs, and generates the following coefficients:

\begin{equation}
\mathbf{W}(k) = \frac{\mathbf{R}_{XX}^{-1}(k)\mathbf{A}(k)}{\mathbf{A}^H(k)\mathbf{R}_{XX}^{-1}(k)\mathbf{A}(k)}
\end{equation}

Finally, the Generalized Eigenvalue Decomposition (GEV) beamformer \cite{heymann2016neural} extracts the principal component using generalized eigenvalue decomposition using the speech and noise SCMs to generate the coefficients:

\begin{equation}
    \mathbf{R}_{SS}(k)\mathbf{W}(k) = \lambda\mathbf{R}_{NN}(k)\mathbf{W}(k)
\end{equation}

Speech enhancement using beamforming methods is appealing for speech recognition as it improves the signal-to-distortion ratio (SDR) without introducing nonlinearities that might hurt the speech recognition performance \cite{du2014robust}.

\subsubsection{Spoken language understanding}

SpeechBrain has several recipes for spoken language understanding (SLU). The SLU recipes demonstrate many useful capabilities of the toolkit, like combining pre-trained tokenizers, language models, and ASR models from other recipes, and using different input sources (audio or text) depending on whether the model is training or testing.

There are currently recipes for three open-source SLU datasets with different levels of complexity: Fluent Speech Commands (FSC) \cite{fluent}, Timers and Such \cite{timers-and-such}, and SLURP \cite{slurp}. The recipes all use attention-based RNN sequence-to-sequence models \cite{bahdanau2014neural} to map the input (either the speech signal or a transcript) to the output (a semantic dictionary containing the intent/slots/slot values for the utterance, as a sequence of characters).

The recipes implement both ``conventional'' SLU (training on ground-truth transcripts) and ``end-to-end'' SLU (training on audio). The conventional ``decoupled'' recipe uses the LibriSpeech ASR model described in \S~\ref{librispeech_section} to transcribe the input signal at test time, instead of using the true transcript. The ``multistage'' \cite{Haghani2018} end-to-end recipe uses the same ASR model but during both training and testing. The ``direct'' \cite{Serdyuk2018} recipe uses a single model to map audio directly to semantics without an intermediate search step. For the ASR-based models, either the default LibriSpeech language model or a language model trained on the SLU dataset transcripts can be used. 

The test accuracy for our FSC recipe is 99.60\%, which is close to the recent SotA CTI model based on wav2vec 2.0 (99.7\% in \cite{seo2021integration}). No comparisons for Timers and Such with other papers are available yet, as the dataset was only released recently \cite{timers-and-such}. 
The performance metrics for SLURP with audio as input are given in Table \ref{table:slurp_audio}. 
Our direct recipe using a wav2vec 2.0 encoder outperforms the HerMiT baseline provided in the original SLURP paper \cite{slurp} across all metrics. 
The recipe is slightly worse than SotA performance achieved by CTI for the intent accuracy metric and closely matches the SLU-F1 metric reported in \cite{seo2021integration}. 
Note that unlike CTI, our recipe currently does not use NLU pre-training and does not take advantage of an application-specific CRF architecture or word-aligned slot and slot value labels; instead, the recipe uses a very simple seq2seq model to predict the semantic dictionary directly. When this seq2seq model is applied directly to the ground-truth transcripts instead of audio, we achieve state-of-the-art results (Table \ref{table:slurp_text}).

\begin{table} 
  \caption{Performance on SLURP (audio as input).}
  \centering
  \small
    \begin{tabular}{p{26mm}p{15mm}p{15mm}p{13mm}p{13mm}p{12mm}p{12mm}} \toprule
        \textbf{Model} & \textbf{\lstinline{scenario}} (accuracy) & \textbf{\lstinline{action}} \newline (accuracy) &	\textbf{\lstinline{intent}} \newline (accuracy) & \textbf{Word-F1} &	\textbf{Char-F1} & \textbf{SLU-F1} \\ \midrule
   HerMiT \cite{slurp} & 85.69 &	81.42 &	78.33 &	69.34 &	72.39 &	70.83\\
   CTI \cite{seo2021integration} & --- & --- & \textbf{86.92} & --- & ---  &  \textbf{74.66} \\ \midrule
   SpeechBrain Direct (CRDNN) & 82.15 & 77.79 & 75.64 & 62.35 & 66.45 & 64.34 \\ 
   SpeechBrain Direct (wav2vec 2.0) & \textbf{89.49} & \textbf{86.40} & 85.34 & \textbf{72.60} & \textbf{76.76} & \textbf{74.62} \\
   \midrule
    \end{tabular}
    \label{table:slurp_audio}
\end{table}

\begin{table}[h!]
  \caption{Performance on SLURP (NLU / ground-truth transcripts as input).}
  \centering
  \small
    \begin{tabular}{p{25mm}p{15mm}p{15mm}p{15mm}} \toprule
        \textbf{Model} & \textbf{\lstinline{scenario}} \newline (accuracy) & \textbf{\lstinline{action}}\newline (accuracy) &	\textbf{\lstinline{intent}}\newline (accuracy) \\ \midrule
   HerMiT \cite{slurp} & 90.15 & 86.99 & 84.84\\ 
   CTI \cite{seo2021integration} & --- & --- & 87.73\\ \midrule
   SpeechBrain NLU & \textbf{91.45} & \textbf{89.46} & \textbf{88.68} \\ 
   \midrule
    \end{tabular}
    \label{table:slurp_text}
\end{table}

\subsection{Architecture details}
In this section, we provide more details on the SpeechBrain architecture outlined in \S~\ref{sec:sb_arch}.
\label{app:arch_det}
\subsubsection{Data preparation}\label{app:data_prep}
The goal of data preparation is to parse a dataset and create the data-manifest files, which contain meta-information about the input data (e.g., file path, annotation, etc.). 
SpeechBrain data-io supports the CSV and JSON file formats, or the user can simply provide a dict. Listing~\ref{lst:manifest-json} reports an excerpt of a JSON data-manifest file for speech recognition:
\begin{lstlisting}[language=json, firstnumber=1, basicstyle=\footnotesize\ttfamily, caption=An excerpt of a JSON data-manifest file for speech recognition., label={lst:manifest-json}]
{
  "sentence001": {
    "wav": "{data_root}/file_snt001.wav",
    "length": 2.10,
    "words": "SWITCH OFF THE LIGHT"
  },
}
\end{lstlisting}
We use a dict (key-value map) structure where each example or spoken utterance is identified and addressable by a unique key or \textit{example ID}.
The entries in each example vary by task and dataset.
For example, in speech recognition, audio files and the corresponding text are needed, whereas, in source separation, we would expect each example to contain the entries for \textit{mixture} and \textit{sources} signals.
The CSV format can also be used:

\clearpage
\begin{lstlisting}[language=json, firstnumber=1, basicstyle=\footnotesize\ttfamily, caption=An excerpt of a CSV data-manifest file for speech recognition.]
ID,length,wav,words
sentence001,2.10,{data_root}/file_snt001.wav,"SWITCH OFF THE LIGHT"
sentence002,2.70,{data_root}/file_snt002.wav,"SWITCH ON THE LIGHT"
sentence003,3.20,{data_root}/file_snt003.wav,"PLEASE, TURN OFF THE LIGHT"
\end{lstlisting}

Dataset parsing scripts, which create the data-manifest files, are provided for many commonly-used speech datasets. Since the data manifests can generally be made relative to the data directory root, data-manifest files can even be provided for download directly, skipping the dataset parsing. All datasets and tasks tend to have at least small subtle differences in formats, and thus SpeechBrain does not have any required entries besides the example ID.

\subsubsection{HyperPyYAML details}
Our primary additions to the YAML format are addition of the following special tags, which are are added before an item definition, and are prefixed with \colorbox{backcolour}{\textbf{\lstinline{!}}}:

\begin{itemize}
    \item \colorbox{backcolour}{\textbf{\lstinline{!new}}}:  instantiates python objects. The object is created with the arguments passed with a list for positional arguments or as a dictionary for keyword arguments.
    \item \colorbox{backcolour}{\textbf{\lstinline{!name}}}: creates function objects. Behind the scenes, it uses functools.partial to create a new function definition with the default arguments provided.
    \item \colorbox{backcolour}{\textbf{\lstinline{!ref}}}: used for referring to a previously-defined item. It can support simple arithmetic and string concatenation for basic hyperparameter combinations.
    \item \colorbox{backcolour}{\textbf{\lstinline{!copy}}}: used to perform a deep copy of a previosly define object. 
    \item \colorbox{backcolour}{\textbf{\lstinline{!tuple}}}: creates tuples.
    \item \colorbox{backcolour}{\textbf{\lstinline{!include}}}: used to insert other YAML files.
    \item \colorbox{backcolour}{\textbf{\lstinline{!apply}}}: loads and executes a python function, storing the result.
\end{itemize}

\subsubsection{Data-io details}

SpeechBrain data-io is built to extend PyTorch \colorbox{backcolour}{\lstinline{data.utils}} and provides the user with several abstractions for reading, encoding, padding and batching data. It is designed with speech processing in mind specifically, but most of the problems it solves are general to variable-length sequence processing.

\begin{figure}[h]
\centering
\includegraphics[width=13cm]{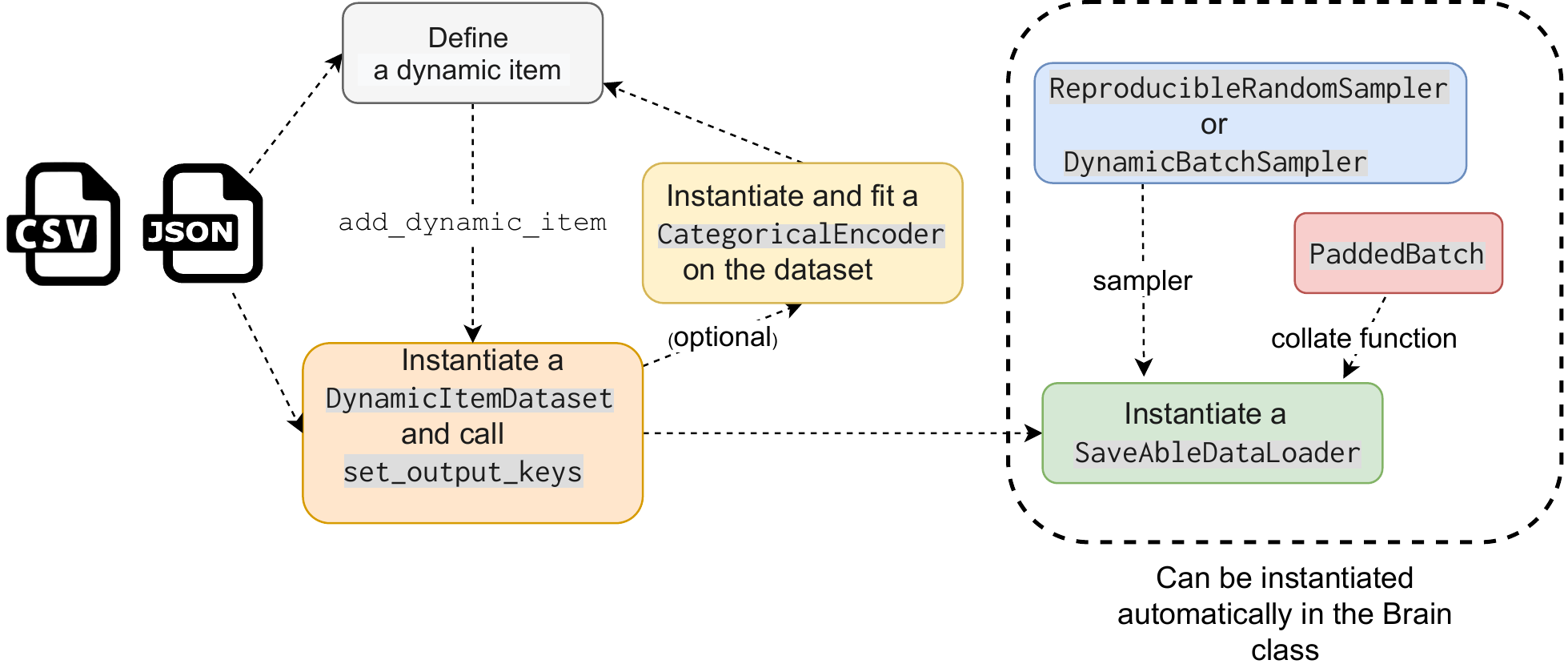}
\caption{An overview of SpeechBrain data-io.}
\label{fig:sbdataio}
\end{figure}

Most of the data-io is built around four abstractions: \colorbox{backcolour}{\lstinline{DynamicItemDataset}}, \colorbox{backcolour}{\lstinline{DynamicItem}}, \colorbox{backcolour}{\lstinline{PaddedBatch}} and \colorbox{backcolour}{\lstinline{SaveableDataloader}}. 
SpeechBrain also provides a \colorbox{backcolour}{\lstinline{CategoricalEncoder}} class which implements label encoding for classification tasks such as speaker recognition. 
The SpeechBrain data-io is illustrated in Figure~\ref{fig:sbdataio}.

Based on a data-manifest (file or dict), a \colorbox{backcolour}{\lstinline{DynamicItemDataset}} can be created. 

\begin{lstlisting}[language=Python, frame=lines, basicstyle=\footnotesize\ttfamily, caption=\colorbox{backcolour}{\lstinline{DynamicItemDataset}} instantiation.]
from speechbrain.dataio.dataset import DynamicItemDataset
train_dataset = DynamicItemDataset.from_json("train.json")
val_dataset = DynamicItemDataset.from_json("val.json")
\end{lstlisting}

Each entry in an example is an \textit{item}, following the Python dict terminology. The items that the data-manifest provides statically are called \textit{static items}: they are kept in memory and stay unchanged. \textit{Dynamic items} are specified by a transformation (e.g., a function) of any number of existing items. These dynamic items are evaluated on-demand. A clear case is a dynamic item that takes a path to an audio file and provides the loaded audio signal. Dynamic items can take other dynamic items as inputs, and a dependency graph is used to determine an evaluation order. Thus, another dynamic item could take the loaded audio signal and provide an augmented version.
A \colorbox{backcolour}{\lstinline{GeneratorDynamicItem}} takes any number of inputs and provides a chain of related dynamic items via the Python generator function syntax. Listings~\ref{lst:read-file-dyn-item}~and~\ref{lst:augment-dyn-item} show implementations of the aforementioned examples: first, a dynamic item loads an audio file, and then a chain of dynamic items augment the loaded signal.
\begin{lstlisting}[language=Python, frame=lines, basicstyle=\footnotesize\ttfamily, caption={A dynamic item which loads an audio file.}, label={lst:read-file-dyn-item}]
@speechbrain.utils.data_pipeline.takes("file_path")
@speechbrain.utils.data_pipeline.provides("signal")
def audio_input(file_path):
    sig = speechbrain.dataio.dataio.read_audio(file_path)
    return sig
\end{lstlisting}

\begin{lstlisting}[language=Python, frame=lines, basicstyle=\footnotesize\ttfamily, caption={Example of a chain of dynamic items, which augments the output of another dynamic item.}, label={lst:augment-dyn-item}]
import random
speechbrain.utils.data_pipeline.takes("sig")
@speechbrain.utils.data_pipeline.provides("rgain", "rgain_offset")
def augmentation(sig):
    random_gain_sig = sig*random.rand()
    yield random_gain_sig
    sig_with_offset = random_gain_sig + 1
    yield sig_with_offset
\end{lstlisting}
The user specifies which items should be returned by the DynamicItemDataset. 
Items are evaluated lazily: only the strictly necessary operations for the user requested items are performed. This allows for significant computational savings and faster execution. 
For example in listing~\ref{lst:set-output-keys} only the example ID (\textit{id}) and speaker ID (\textit{spkid}) static items and the random gain augmentation dynamic item (\textit{rgain}) are requested.

\begin{lstlisting}[language=Python, frame=lines, basicstyle=\footnotesize\ttfamily, caption={Setting output items for the train\_dataset \colorbox{backcolour}{\lstinline{DynamicItemDataset}}}, label={lst:set-output-keys}]
speechbrain.dataio.dataset.set_output_keys(
        [train_dataset], ["id, "spkid", "rgain"],
    )
\end{lstlisting}

Since the audio tensor with offset (\textit{rgain\_offset}) is not requested it is not computed at all in this example. On the contrary the audio tensor \textit{sig} is needed for computing \textit{rgain} and thus it is evaluated. 

The \colorbox{backcolour}{\lstinline{DynamicItemDataset}} can thus provide multiple different views of the same dataset on demand. Iterating over the dataset can be extremely fast if the user only needs a particular item, e.g., to fit a \colorbox{backcolour}{\lstinline{CategoricalEncoder}}. 
Listing~\ref{lst:create-cat-enc} shows fitting a \colorbox{backcolour}{\lstinline{CategoricalEncoder}} in a speaker identification task, continuing the above examples.

\clearpage
\begin{lstlisting}[language=Python, frame=lines, basicstyle=\footnotesize\ttfamily, caption={Fitting a \colorbox{backcolour}{\lstinline{CategoricalEncoder}} for speaker recognition. This only evaluates the \textit{spkid} item.}, label={lst:create-cat-enc}]
from speechbrain.dataio.encoder import CategoricalEncoder
spk_id_encoder = CategoricalEncoder()
spk_id_encoder.update_from_didataset(dataset, "spkid")
train_dataset.add_dynamic_item(spk_id_encoder.encode_label, takes="spkid", provides="spkid_enc")
\end{lstlisting}
SpeechBrain also provides \colorbox{backcolour}{\lstinline{CategoricalEncoder}} sub-classes for encoding text and handle special tokens for the training of sequence-to-sequence models.

A \colorbox{backcolour}{\lstinline{DynamicItemDataset}} object can be wrapped by a standard PyTorch \colorbox{backcolour}{\lstinline{DataLoader}} or by SpeechBrain \colorbox{backcolour}{\lstinline{SaveableDataloader}}. The \colorbox{backcolour}{\lstinline{Brain}} class can handle this automatically for the user and uses the  \colorbox{backcolour}{\lstinline{SaveableDataloader}} by default with \colorbox{backcolour}{\lstinline{PaddedBatch}} as the default \textit{collate function}.
and injecting \colorbox{backcolour}{\lstinline{ReproducibleRandomSampler}} as the \textit{sampler} in case \colorbox{backcolour}{\lstinline{shuffle=True}} .
More in general, custom PyTorch \textit{samplers} and \textit{collate functions} can be integrated seamlessly in SpeechBrain data-io. 

\colorbox{backcolour}{\lstinline{SaveableDataloader}} allows  for intra-epoch checkpointing, a feature that is useful when running extremely computationally demanding experiments where each epoch can take several hours. 

\colorbox{backcolour}{\lstinline{PaddedBatch}} is both a \textit{collate function} and a batch object.
It handles for the user the rather annoying task of padding examples together. 
By default, it batches together only PyTorch tensors by adding zeros to the right on the last dimension. Other data types are not batched together but, instead, are returned in a python list. 
It also provides a semantically meaningful interface, as shown in listing~\ref{lst:padded-batch}.

\begin{lstlisting}[language=Python, frame=lines, basicstyle=\footnotesize\ttfamily, caption={Accessing in \colorbox{backcolour}{\lstinline{PaddedBatch}} each requested item as well as the relative lengths of the padded data}, label={lst:padded-batch}]
from speechbrain.dataio.dataloader import make_dataloader
train_dataset.set_output_keys(["id", "rgain"])
dataloader = make_dataloader(train_dataset, batch_size=8)
for batch in dataloader:
    # Access a list of the example IDs in this batch
    batch.id
    # Access the speech data:
    batch.rgain.data
    # Access the relative lengths:
    batch.rgain.lengths
\end{lstlisting}

\subsubsection{\colorbox{backcolour}{\lstinline{Brain}} class details}\label{sec:brain_api_details}
The \colorbox{backcolour}{\lstinline{Brain}} class implements customizable methods for managing the different aspects of training and evaluation. 
Table \ref{tab:sb_methods} describes more in detail these useful methods.

\begin{table}[t!]
  \caption{Main methods implemented in the \colorbox{backcolour}{\lstinline{Brain}} class.}
  \centering
  \small
    \begin{tabular}{p{28mm}p{105mm}} \toprule
        \textbf{Method} & \textbf{Description} \\ \midrule
   \colorbox{backcolour}{\textbf{\lstinline{fit}}} & Main function for training. It iterates epochs and datasets to improve the objective. \\ \midrule 
   \colorbox{backcolour}{\textbf{\lstinline{fit\_batch}}}  & Trains a batch. It calls \colorbox{backcolour}{\textbf{\lstinline{compute\_forward}}}, \colorbox{backcolour}{\textbf{\lstinline{compute\_objectives}}}, and optimizes the loss. \\ \midrule
   \colorbox{backcolour}{\textbf{\lstinline{compute\_forward}}} & Defines computations from input to output predictions.\\ \midrule
   \colorbox{backcolour}{\textbf{\lstinline{compute\_objective}}} & Defines computations from  predictions to loss.\\ \midrule
   \colorbox{backcolour}{\textbf{\lstinline{on\_stage\_start}}} & Gets called at the beginning of a epoch. Useful for metric initialization.\\ \midrule
   \colorbox{backcolour}{\textbf{\lstinline{on\_stage\_end}}} & Gets called at the end of a epoch. Useful for statistics, checkpointing, learning rate annealing.\\\midrule
    \end{tabular}
\label{tab:sb_methods}
\end{table}

The Brain class only takes the following arguments:
\begin{itemize}
    \item \colorbox{backcolour}{\textbf{\lstinline{modules}}}: takes a dictionary of PyTorch modules and converts it to a PyTorch \colorbox{backcolour}{\lstinline{ModuleDict}}. provides a convenient way to move all parameters to the correct device, call \colorbox{backcolour}{\lstinline{train()}} and \colorbox{backcolour}{\lstinline{eval()}}, and wrap the modules in the appropriate distributed wrapper if necessary.
    \item \colorbox{backcolour}{\textbf{\lstinline{opt\_class}}}:  takes a function definition for a PyTorch optimizer. The reason for choosing this as input rather than a pre-constructed PyTorch optimizer is that the \colorbox{backcolour}{\lstinline{Brain}} class automatically handles wrapping the module parameters in distributed wrappers if requested. That needs to happen before the parameters get passed to the optimizer constructor.
    \item \colorbox{backcolour}{\textbf{\lstinline{hparams}}}:  accepts a dictionary of hyperparameters that will be accessible to all the internal methods.
    \item \colorbox{backcolour}{\textbf{\lstinline{run\_opts}}}: there are a large number of options for controlling the execution details for the \colorbox{backcolour}{\lstinline{fit()}} method, that can all be passed via this argument. Some examples include enabling debug mode, the execution device, and the distributed execution options.
    \item \colorbox{backcolour}{\textbf{\lstinline{checkpointer}}}: it is a pointer to the SpeechBrain checkpointer.
    This way, at the beginning of training, the most recent checkpoint is loaded and training is resumed from that point. If training is finished, this moves on to evaluation. During training, the checkpoints are saved every 15 minutes by default. At the beginning of the evaluation, the "best" checkpoint is loaded, as determined by the lowest or highest score on a metric recorded in the checkpoints.
\end{itemize}

\subsubsection{Lobes}
In neuroscience, the lobes are areas of the brain associated with some specific high-level functionality. Similarly, in SpeechBrain we collect common higher-level speech processing functionalities in the lobe folder. For instance, lobes contain popular models used for speech processing, as reported in Table \ref{tab:lobes_models}. Moreover, we implement here data augmentation strategies, as discussed in Table \ref{tab:lobes_aug}.

\begin{table}[h!]
  \caption{Main models implemented in lobes.}
  \centering
  \small
    \begin{tabular}{p{20mm}p{26mm}p{95mm}} \toprule
        \textbf{Method} & \textbf{Main use} & \textbf{Description} \\ \midrule
   CRDNN & Speech recognition & A combination of convolutional, recurrent, and fully-connected networks. Layer and batch normalization are used for feedforward layers. Time-pooling can be optionally used for downsampling. Users can select the type of RNN to plug in (e.g, LSTM \cite{lstm}, GRU \cite{gru2}, LiGRU \cite{li_gru}, vanilla RNN). \\ \midrule 
   TransformerASR & Speech recognition & A basic sequence-to-sequence transformer \cite{transformers} for speech recognition. \\ 
   \midrule 
   ECAPA-TDNN & Speaker recognition & The ECAPA-TDNN model \cite{ecapa} employs a channel- and context-dependent attention mechanism, Multilayer Feature Aggregation (MFA), as well as Squeeze-Excitation (SE) and residual
blocks. \\ \midrule 
   X-vector & Speaker recognition & Employs a TDNN \cite{Lang+Hinton88,waibel} followed by a statistical pooling layer. It is used to compute x-vector embeddings \cite{xvector}. \\ \midrule 
   MetricGAN & Speech enhancement & Implements a LSTM-based generator followed by a discriminator that estimates the quality of speech using PESQ. \\ \midrule 
   ConvTasNet & Speech separation & Uses a linear encoder to generate a representation of the speech waveform. Speaker separation is achieved by applying a mask to the encoded representation. The mask encoded representations are then converted back to the waveforms using a linear decoder \cite{convtasnet}. \\ \midrule 
   Dual-Path & Speech separation & Splits long speech inputs into smaller chunks and applies intra- and inter-chunk operations over them \cite{luo2020dualpath}. \\ \midrule 
   SepFormer & Speech separation & Couples the Dual-Path framework with an efficient multi-scale transformers approach \cite{sepformer}. \\ \midrule 
    \end{tabular}
    \label{tab:lobes_models}
\end{table}

\begin{table}[h!]
  \caption{Data augmentation techniques implemented in lobes.}
  \centering
  \small
    \begin{tabular}{p{40mm}p{100mm}} \toprule
        \textbf{Method} & \textbf{Description} \\ \midrule
   SpecAugment & It applies time and frequency masking as well as time warping to the input spectrum (frequency-domain implementation) \cite{Park2019}. \\ \midrule 
   Time-Domain SpecAugment & It applies time/frequency masking and time warping to the input waveform (time-domain implementation). Each disturbance is randomly activated according to the specified activation probabilities. \\ \midrule 
   Environmental Corruption & It adds noise, reverberation, and bubble (i.e., noisy from many speakers talking in the background).  Each corruption technique is randomly activated according to the specified activation probabilities. The amount of noise added is controlled with proper settings.
   When not specified directly, we use the noise and impulse responses from the OpenRIR dataset\footnote{\url{http://www.openslr.org/28/}}. 
   \\ \midrule
    \end{tabular}
    \label{tab:lobes_aug}
\end{table}

\subsubsection{Inference}
To make inference with pre-trained models easier, we provide some inference classes able to support a variety of speech tasks.
For instance, it is possible to transcribe an input sentence using a speech recognizer with just a few lines of code:
\begin{lstlisting}[language=Python, frame=lines, caption=Inference with a speech recognizer.]
from speechbrain.pretrained import EncoderDecoderASR

asr_model = EncoderDecoderASR.from_hparams(
    source="speechbrain/asr-transformer-transformerlm-librispeech", savedir="pretrained_models/asr")
asr_model.transcribe_file("example.wav")
>>> ["THE BIRCH CANOE SLID ON THE SMOOTH PLANKS"]
\end{lstlisting}
The inference API relies on a YAML similar to that used for training. 
Another example for speaker verification is reported in the following:
\begin{lstlisting}[language=Python, frame=lines, caption=Speaker verification inference.]
from speechbrain.pretrained import SpeakerRecognition
file1= "speechbrain/spkrec-ecapa-voxceleb/example1.wav"
file2= "speechbrain/spkrec-ecapa-voxceleb/example2.wav"
verification = SpeakerRecognition.from_hparams(
    source="speechbrain/spkrec-ecapa-voxcebeb", 
    savedir="pretrained_models/spkrec-ecapa-voxceleb")
score, prediction = verification.verify_files(file1,file2)
\end{lstlisting}
In this case, we feed the verification system with two audio files, and the outcome is "0" if the files are from different speakers and "1" otherwise. 
We have shared our best-performing models on the Hugging Face hub\footnote{\url{huggingface.co/speechbrain}}.

\subsection{Experiment details}
\label{sec:exp_details}
In the following, we provide more details on the datasets, evaluation metrics, and experimental settings used in the experiments reported in the paper.
\label{sec:details}
\subsubsection{Datasets}
As shown in Table \ref{tab:tasks}, SpeechBrain already provides recipes for several common speech corpora\footnote{The datasets used for our research are anonymized and do not contain personally identifiable
information or offensive content.
Datasets available through LDC require that participants consented to share their data in a corpus. Unless explicitly mentioned, we were not able to find the consent information for the other datasets. However, we only use popular corpora, and we have reason to believe that creators explicitly asked for consent from the contributors. 
}:

\begin{itemize}
    \item \textbf{TIMIT} \cite{timit}: The TIMIT corpus contains about 5 hours of speech from 630 speakers of eight major dialects of American English, each reading ten phonetically rich sentences. It includes audio signals sampled at  16kHz (16-bit) resolution and the phonetic transcription of each sentence using the SAMPA phoneme set. TIMIT is licensed by the Linguistic Data Consortium (LDC).
    \item \textbf{LibriSpeech} \cite{librispeech}: LibriSpeech is a corpus of approximately 1000 hours of 16kHz read English speech. The data is derived from audiobooks from the LibriVox project\footnote{\url{https://librivox.org/}}. The volunteers gave their consent to donated their recordings to the public domain. The training data is split into three partitions of 100hr, 360hr, and 500hr sets while the dev and test data are split into the `clean’ and `other’ categories, respectively. Each of the dev and test sets is around 5hr. The corpus is publicly available with the Creative Commons Attribution 4.0 License. 
    \item \textbf{Common Voice} \cite{commonvoice:2020}: Common Voice is Mozilla's initiative to create a free database for speech recognition software. The project is supported by volunteers who record sample sentences with a microphone and review recordings of other users. The website clearly informs the volunteers of the purpose of the recordings.
    The text is derived from different open-source text sources, including Wikipedia. As of May 2021, the dataset contains 7.3k hours of transcribed and validated speech in 60 languages. 
Our paper uses the latest released version of the corpus (Common Voice 6.1). The dataset is publicly available with the Creative Commons Attribution 4.0 License. 
    \item \textbf{VoxCeleb} \cite{voxceleb2}: VoxCeleb is an audio-visual dataset consisting of short clips of human speech, extracted from interview videos uploaded to YouTube. In this paper, we used both VoxCeleb1 \cite{voxceleb} and voxceleb2 \cite{voxceleb2}. VoxCeleb1 contains over 100,000 utterances for 1,251 celebrities. VoxCeleb2 contains over a million utterances for 6,112 identities. The dataset is available to download under a Creative Commons Attribution 4.0 International License.
    \item \textbf{AMI} \cite{ami-corpus}: The AMI Meeting Corpus is a widely used multi-modal dataset consisting of 100 hours of meeting recordings. The meetings have been recorded with both close-talking and far-field microphones that are time-synchronized. The meetings are majorly divided into a scenario and non-scenario meetings. In a scenario meeting, four participants play a specific role assigned to them. The non-scenario ones, instead, include a general discussion between three to four participants. The AMI dataset also has fixed official splits for various tasks to foster replicability. The signals, transcription, and annotations, have been released publicly under the Creative Commons Attribution 4.0 International Licence (CC BY 4.0).
    \item \textbf{Voicebank-DEMAND} \cite{valentini2017noisy}:
    It contains speech of 30 clean speakers extracted from the Voice Bank corpus \cite{voicebank}: 28 are included in the training set, and two are in the validation set. The noisy speech is synthesized by contaminating the clean signals with noise from Diverse Environments Multichannel Acoustic Noise Database (DEMAND) \cite{demand}. Both speakers and noise conditions in the test set are unseen by the training set. The training and test set contains 11572 and 824 noisy-clean speech pairs, respectively. The dataset is available to the community with the Creative Commons Attribution 4.0 International Public License.
    \item \textbf{WSJ0-mix} \cite{deepclustering}: It is a single-channel speech separation dataset derived from the Wall Street Journal corpus (licensed by LDC). It contains mixtures of two or three speakers. The training set consists of 30 hours of overlapped speech material that was generated by randomly selecting utterances by different speakers from the WSJ0 training set \textit{si\_tr\_s}, and by mixing them at various signal-to-noise ratios (SNR).
\end{itemize}

\subsubsection{Evaluation metrics}
SpeechBrain supports all the standard evaluation metrics needed to assess the performance of the proposed tasks.
In the following, we report a short description of the  evaluation metrics used in this paper:

\begin{itemize}
    \item \textbf{Word Error Rate (WER\%)}:
    The WER(\%) is derived from the Levenshtein distance and compares a reference and a hypothesized word-level transcription. It is computed by summing up the number of word insertions, deletions, substitutions and dividing it by the total number of words in the reference transcription.  
    Listing \ref{lst:wer1} shows an example of the WER summary provided by SpeechBrain, where the alignment between the reference and the hypothesized transcription are provided as well. 
    \item \textbf{Phone Error Rate (PER\%)}:
    It is the same as the WER, but it is computed using phonemes as basic units rather than words.
    \item \textbf{Equal Error Rate (EER\%)}:
    It corresponds to the error rate achieved when the false acceptance rate and the false rejection rate are equal. The lower the EER is, the higher is the accuracy of the system.
    \item \textbf{Diarization Error Rate (DER\%)}:
Diarization error rate (DER) is the standard metric for evaluating speaker diarization systems. It is defined as:
\begin{equation}
    DER = \frac{false\;alarm + missed + confusion}{reference\;length}
\end{equation}
where \textit{false alarm} is the length of non-speech incorrectly classified as speech, and \textit{missed} detection
is the length of segments that are considered as speech in reference, but not in hypothesis. \textit{confusion} is the length of segments that are assigned to different speakers in hypothesis and reference, while \textit{reference-length} is the total duration of speech in the reference. The lower DER is, the better the diarization system is.
\item \textbf{Perceptual Evaluation of Speech Quality (PESQ)}:
It is a complex metric designed to predict subjective opinion scores of a degraded audio sample \cite{pesq}. 
PESQ (full reference modality) compares the clean and noisy signals and returns a score from 4.5 to -0.5, with higher scores indicating better quality.

\item \textbf{MOS predictor of overall signal quality (COVL)}: The COVL metric is part of a set of three common metrics of enhancement quality, along with CSIG and CBAK. These metrics are a composite of other commonly used metrics, like PESQ and Itakura-Saito distance measure. The resulting metric showed a much higher correlation with human judgments than any contributing metric~\cite{hu2007evaluation}.

\item \textbf{Scale-invariant signal-to-noise ratio improvement (SI-SNRi)}: 
Scale-invariant signal-to-noise ratio improvement (SI-SNRi) is a performance metric for source separation \cite{convtasnet}, proposed as an alternative to the Source-to-Distortion Ratio \cite{bsseval}. It is defined as follows: 

\begin{align*}
    s_\textbf{target} :=& \frac{(\widehat s^\top s)s }{\| s\|^2},  \\
    e_\textbf{noise} :=& \widehat s - s_\textbf{target}, \\
    \text{SI-SNR} :=& 10 \log_{10} \frac{ \| s_\textbf{target} \|^2} { \| e_\textbf{noise} \|^2},
\end{align*}
where $s \in \mathbb R^T$ is the ground truth source, and $\widehat s \in \mathbb R^T$ is the source estimated by the model, and $\| s \|^2 = s^\top s$, denotes the $l_2$ norm operation. The scale-invariance is ensured by removing the mean from $s$ and $\widehat s$, and dividing them by their respective standard deviations before calculating the SNR. Finally, SI-SNR improvement, (SI-SNRi) is calculated as follows:

\begin{align*}
    \text{SI-SNRi} := \text{SI-SNR}(s, \widehat{s}) - \text{SI-SNR}(s, x),
\end{align*}
where $x \in \mathbb R^T$ denotes the mixture signal corresponding to the source $s$.
\end{itemize}

\clearpage
\begin{lstlisting}[label=lst:wer1, frame=lines, caption=Excerpt of the summary file generated by SpeechBrain for the WER metric described above.]
%WER 2.46 [ 1291 / 52576, 169 ins, 124 del, 998 sub ]
%SER 28.55 [ 748 / 2620 ]
Scored 2620 sentences, 0 not present in hyp.
=========================================
ALIGNMENTS

Format:
<utterance-id>, WER DETAILS
<eps> ; reference  ; on ; the ; first ;  line
  I   ;     S      ; =  ;  =  ;   S   ;   D  
 and  ; hypothesis ; on ; the ; third ; <eps>
=========================================
61-70968-0058, %WER 0.00 [ 0 / 5, 0 ins, 0 del, 0 sub ]
WILL ; YOU ; FORGIVE ; ME ; NOW
 =   ;  =  ;    =    ; =  ;  = 
WILL ; YOU ; FORGIVE ; ME ; NOW
=========================================
5142-33396-0000, %WER 20.00 [ 1 / 5, 0 ins, 0 del, 1 sub ]
AT ; ANOTHER ; TIME ; HARALD ; ASKED
=  ;    =    ;  =   ;   S    ;   =  
AT ; ANOTHER ; TIME ; HAROLD ; ASKED
=========================================
237-134500-0005, %WER 11.11 [ 1 / 9, 0 ins, 1 del, 0 sub ]
OH ; BUT ; I'M ; GLAD ; TO ; GET ; THIS ; PLACE ; MOWED
=  ;  =  ;  =  ;  =   ; =  ;  =  ;  =   ;   =   ;   D  
OH ; BUT ; I'M ; GLAD ; TO ; GET ; THIS ; PLACE ; <eps>
=========================================
260-123288-0012, %WER 14.29 [ 1 / 7, 1 ins, 0 del, 0 sub ]
THAT ; WILL ; BE ; <eps> ; SAFEST ; NO ; NO ; NEVER
 =   ;  =   ; =  ;   I   ;   =    ; =  ; =  ;   =  
THAT ; WILL ; BE ;  THE  ; SAFEST ; NO ; NO ; NEVER
\end{lstlisting}

\subsubsection{Experimental setups}
In this section, we report more details for the experiments reported in the paper. 
For lower-level detail, please refer to the project repository directly\footnote{\url{github.com/speechbrain/speechbrain}}.
The hyperparameters of the models were initially based on values reported in the literature for similar models. Then, several experiments were carried out to progressively derive better hyperparameters.
We use 32GB NVIDIA V100 in our experiments.
The best hyperparameters found so far are summarized in the following tables.

\begin{table}[h!]
  \caption{Main hyperparameters used in the reported LibriSpeech experiments.}
  \centering
  \small
    \begin{tabular}{p{25mm}p{15mm}p{15mm}p{70mm}} \toprule
        \textbf{Task} & \textbf{Dataset} & \textbf{Technique} &
        \textbf{Experimental Setting}
        \\ \midrule
   Speech recognition & LibriSpeech & CTC+Att (RNN) & Encoder: CRDNN (2 CNNs, 4 LSTM, 1 DNN layers) \newline
   Decoder: GRU (1 layer) + Beam search + LM \newline
   Augmentation: yes \newline
   Features: 40 fbanks \newline
   Pretraining: no \newline
   Dropout: 0.15 (for both encoder and decoder) \newline
   Batchnorm: yes \newline
   Number of epochs: 25 \newline
   Batch size: 8 \newline
   Learning rate: 1.0 \newline
   LR scheduler: new bob \newline
   Optimizer: Adam \newline
   Loss: CTC+NLL \newline
   CTC weight: 0.5 \newline
   Number of tokens: 5000 \newline
   Training Time: 5h 20m/epoch (on a NVIDIA V100) 
   \\ \midrule
   Speech recognition & LibriSpeech &
   CTC+Att (Transf.) & Encoder: ContextNet (3 lay) + Transformer (12 lay) \newline
   Decoder: Transformer (6 layers) + Beam search + LM \newline
   Augmentation: yes \newline
   Features: 80 fbanks \newline
   Pretraining: no \newline
   Dropout: no (for both encoder and decoder) \newline
   Layernorm: yes \newline
   Number of epochs: 110 \newline
   Batch size: 16 \newline
   Gradient accumulation: 4 \newline
   Gradient clipping: 5.0 \newline
   Learning rate: 1.0 \newline
   Learning rate (fine tune with SGD): 0.000025 \newline
   LR scheduler: new bob \newline
   Optimizer: Adam \newline
   Loss: CTC+NLL \newline
   CTC weight: 0.4 \newline
   Number of tokens: 5000 \newline
   Training Time: 1h 50m/epoch (on 2 NVIDIA V100)
   \\ \midrule
   \end{tabular}
\end{table}

\begin{table}[h!]
  \caption{Main hyperparameters used in the reported TIMIT experiments.}
  \centering
  \small
    \begin{tabular}{p{25mm}p{15mm}p{15mm}p{70mm}} \toprule
        \textbf{Task} & \textbf{Dataset} & \textbf{Technique} &
        \textbf{Experimental Setting}
        \\ \midrule
   Speech recognition & TIMIT & CTC & Model: CRDNN (2 CNNs, 4 LiGRUs, 2 DNN layers) \newline
   Augmentation: yes \newline
   Features: 40 fbanks \newline
   Pretraining: no \newline
   Dropout: 0.15 (encoder), 0.5 (decoder) \newline
   Batchnorm: yes \newline
   Number of epochs: 50 \newline
   Batch size: 8 \newline
   Learning rate: 1.0 \newline
   LR scheduler: new bob \newline
   Optimizer: Adam \newline
   Loss: CTC \newline
   Training Time: 2m 25sec/epoch (on a NVIDIA V100) 
   \\ \midrule
   Speech recognition & TIMIT & Transducer & Model: CRDNN (2 CNNs, 4 LiGRUs, 2 DNN layers) \newline
   Augmentation: yes \newline
   Features: 40 fbanks \newline
   Pretraining: no \newline
   Dropout: 0.15 (encoder), 0.5 (decoder) \newline
   Batchnorm: yes \newline
   Number of epochs: 50 \newline
   Batch size: 8 \newline
   Learning rate: 1.0 \newline
   LR scheduler: new bob \newline
   Optimizer: Adadelta \newline
   Loss: Transducer Loss \newline
   Training Time: 1m 10 sec/epoch (on a NVIDIA V100) 
   \\ \midrule
   Speech recognition & TIMIT & CTC+Att & Encoder: CRDNN (2 CNNs, 4 LiGRUs, 2 DNN layers) \newline
   Decoder: GRU (1 layer) + Beam search \newline
   Augmentation: yes \newline
   Features: 40 fbanks \newline
   Pretraining: no \newline
   Dropout: 0.15 (encoder), 0.5 (decoder) \newline
   Batchnorm: yes \newline
   Number of epochs: 20 \newline
   Batch size: 8 \newline
   Learning rate: 0.0003 \newline
   LR scheduler: new bob \newline
   Optimizer: Adam \newline
   Loss: CTC+NLL Loss \newline
   CTC weight: 0.2 \newline
   Training Time: 2m 25 sec/epoch (on a NVIDIA V100) 
   \\ \midrule
   Speech recognition & TIMIT & CTC+Att+ SSL & Encoder: wav2vec (Transformer) \newline
   Decoder: GRU (1 layer) + Beam search \newline
   Augmentation: yes \newline
   Features: 40 fbanks \newline
   Pretraining: wav2vec2-large-lv60 (Hugging Face) \newline
   Dropout: 0.1 (encoder), 0.5 (decoder) \newline
   Batchnorm: yes \newline
   Number of epochs: 50 \newline
   Batch size: 8 \newline
   Learning rate: 0.0003 \newline
   Learning rate : 0.0001 \newline
   LR scheduler: new bob \newline
   Optimizer: Adam \newline
   Loss: CTC+NLL Loss \newline
   CTC weight: 0.1 \newline
   Training Time: 3m 14 sec/epoch (on a NVIDIA V100) 
   \\ \midrule
   \end{tabular}
\end{table}

\begin{table}[h!]
  \caption{Main hyperparameters used in the reported Common Voice experiments.}
  \centering
  \small
    \begin{tabular}{p{24mm}p{16mm}p{15mm}p{70mm}} \toprule
        \textbf{Task} & \textbf{Dataset} & \textbf{Technique} &
        \textbf{Experimental Setting}
        \\ \midrule
   Speech recognition & Common Voice & CTC+Att & Encoder: CRDNN (3 CNNs, 5 LSTM, 2 DNN layers) \newline
   Decoder: GRU (1 layer) + Beam search \newline
   Augmentation: yes \newline
   Features: 80 fbanks \newline
   Pretraining: no \newline
   Dropout: 0.15 (for both encoder and decoder) \newline
   Batchnorm: yes \newline
   Number of epochs: 50 \newline
   Batch size: 12 \newline
   Learning rate: 1.0 \newline
   LR scheduler: new bob \newline
   Optimizer: Adadelta \newline
   Loss: CTC+NLL Loss \newline
   Number of tokens: 500 \newline
   CTC weight: 0.3 \newline
   Training Time (En): 6h 40 min/epoch (NVIDIA V100) \newline
   Training Time (Fr): 3h 20 min/epoch (NVIDIA V100) \newline 
   Training Time (It): 1h 00 min/epoch (NVIDIA V100) \newline
   Training Time (Kw): 4h 30 min/epoch (NVIDIA V100) 
   \\ \midrule
   Speech recognition & Common Voice & CTC+Att + SSL & Encoder:  (Transformer) \newline
   Decoder: GRU (1 layer) + Beam search \newline
   Augmentation: yes \newline
   Features: 80 fbanks \newline
   Pretraining (En): 2-large-lv60 \newline
   Pretraining (Fr): wav2vec2-large-100k-voxpopuli \newline
   Pretraining (It): wav2vec2-large-100k-voxpopuli \newline
   Pretraining (Kw): wav2vec2-large-100k-voxpopuli\newline
   Dropout: 0.15 (for decoder) \newline
   Batchnorm: yes \newline
   Number of epochs: 30 \newline
   Batch size: 12 \newline
   Learning rate: 1.0 \newline
   Learning rate wav2vec2: 0.0001 \newline
   LR scheduler: new bob \newline
   Optimizer: Adadelta \newline
   Loss: CTC+NLL Loss \newline
   Number of tokens: 500 \newline
   CTC weight: 0.3 \newline
   Training Time (En): 8h 20 min/epoch (2 NVIDIA V100) \newline
   Training Time (Fr): 4h 05 min/epoch (2 NVIDIA V100) \newline 
   Training Time (It): 1h 30 min/epoch (NVIDIA V100) \newline
   Training Time (Kw): 6h 00 min/epoch (NVIDIA V100) 
   \\ \midrule
   \end{tabular}
\end{table}

\begin{table}[h!]
  \caption{Main hyperparameters used in the Speaker Recognition and Diarization experiments.}
  \centering
  \small
    \begin{tabular}{p{26mm}p{16mm}p{15mm}p{70mm}} \toprule
        \textbf{Task} & \textbf{Dataset} & \textbf{Technique} &
        \textbf{Experimental Setting}
        \\ \midrule
   Speaker recognition & Voxceleb2 & x-vector + PLDA & Model:  x-vector (5 TDNN layers) + statistical pool + MLP class\newline
   Augmentation: yes \newline
   Features: 80 fbanks \newline
   Pretraining: no \newline
   Dropout: no \newline
   Batchnorm: yes \newline
   Number of epochs: 20 \newline
   Batch size: 256 \newline
   Learning rate initial: 0.001 \newline
   Learning rate final: 0.0001 \newline
   LR scheduler: linear decay \newline
   Optimizer: Adam \newline
   Loss: NLL Loss \newline
   Training Time (vox1+vox2): 4h 20 min/epoch (NVIDIA V100)
   \\ \midrule
   Speaker recognition & Voxceleb2 & ECAPA-TDNN + cosine dist & Model:  ECAPA-TDNN (5 tdnn layers) + att pooling + MLP class\newline
   Augmentation: yes \newline
   Features: 80 fbanks \newline
   Pretraining: no \newline
   Dropout: no \newline
   Batchnorm: yes \newline
   Number of epochs: 12 \newline
   Batch size: 32 \newline
   Learning base: 0.00000001 \newline
   Learning rate max: 0.0001 \newline
   LR scheduler: CyclicLRScheduler \newline
   Optimizer: Adam \newline
   Loss: NLL Loss \newline
   Training Time (vox1+vox2): 12h 10 min/epoch (NVIDIA V100)
   \\ \midrule
   Speaker diarization & AMI & ECAPA-TDNN + spectral clustering & Embeddings:  ECAPA-TDNN \newline
   Clusteting: Spectral Clustering \newline
   Split type: full\_corpus\_asr \newline
   Skip\_TNO: True \newline
   Mic type: BeamformIt \newline
   VAD type: oracle \newline
   Max subseg dur: 3.0 \newline
   Overlap: 1.5 \newline
   Affinity: cos \newline
   Max num spkrs: 10  \newline
   Oracle \# spkrs: True \newline
   Ignore overlap: True \newline
   Forgiveness collar: 0.25 
   \\ \midrule
   \end{tabular}
\end{table}

\begin{table}[h!]
  \caption{Main hyperparameters used for the speech enhancement experiments.}
  \centering
  \small
    \begin{tabular}{p{26mm}p{16mm}p{15mm}p{70mm}} \toprule
        \textbf{Task} & \textbf{Dataset} & \textbf{Technique} &
        \textbf{Experimental Setting}
        \\ \midrule
   Speech enhancement & Voicebank-DEMAND & MimicLoss & Enhanced Model:  CNN + Transformer \newline
   ASR Model:  CRDNN \newline
   Features: Spectrogram \newline
   Pretraining: no \newline
   Dropout: 0.15 (CRDNN) \newline
   Batchnorm: yes \newline
   Number of epochs: 20 \newline
   Batch size: 256 \newline
   Learning rate: 0.0001 \newline
   Optimizer: Adam \newline
   Loss: NLL+MSE Loss \newline
   Training Time (enhance): 26 min/epoch (NVIDIA V100) \newline
   Training Time (perceptual): 11 min/epoch (NVIDIA V100) \newline
   Training Time (ASR): 5 min/epoch (NVIDIA V100) 
   \\ \midrule
   Speech enhancement & Voicebank-DEMAND & MetricGAN+ & Enhanced Model:  LSTM (2 layers) \newline
   Discriminator Model:  CNN (3 layers) + DNN (3 layers) \newline
   Features: STFT \newline
   Pretraining: no \newline
   Batchnorm: yes \newline
   Number of epochs: 600 \newline
   Batch size: 256 \newline
   Learning rate: 0.0005 \newline
   Optimizer: Adam \newline
   Loss: MSE + PESQ Loss \newline
   Training Time: 11 min/epoch (NVIDIA V100)
   \\ \midrule
   \end{tabular}
\end{table}

\begin{table}[h!]
  \caption{Main hyperparameters used in the speech separation experiments.}
  \centering
  \small
    \begin{tabular}{p{24mm}p{16mm}p{15mm}p{70mm}} \toprule
        \textbf{Task} & \textbf{Dataset} & \textbf{Technique} &
        \textbf{Experimental Setting}
        \\ \midrule
   Speech separation & WSJ0-MIX & ConvTasNET & Model:  ConvTasNET (Encoder, MaskNET, Decoder)\newline
   Augmentation: yes \newline
   Features: waveform \newline
   Pretraining: no \newline
   Dropout: no \newline
   Normalization: GlobalLayerNorm \newline
   Number of epochs: 200 \newline
   Batch size: 1 \newline
   Learning rate: 0.00015 \newline
   LR scheduler: ReduceLROnPlateau \newline
   Optimizer: Adam \newline
   Loss: si-snr with pit-wrapper \newline
   Training Time: 1h 00 min/epoch (NVIDIA V100)
   \\ \midrule
   Speech separation & WSJ0-MIX & DualPathRNN & Model:  DualPathRNN (Encoder, MaskNET, inter-intra RNNs, Decoder)\newline
   Augmentation: yes \newline
   Features: waveform \newline
   Pretraining: no \newline
   Dropout: no \newline
   Normalization:  GlobalLayerNorm \newline
   Number of epochs: 200 \newline
   Batch size: 1 \newline
   Learning rate: 0.00015 \newline
   LR scheduler: ReduceLROnPlateau \newline
   Optimizer: Adam \newline
   Loss: si-snr with pit-wrapper \newline
   Training Time: 3h 00 min/epoch (NVIDIA V100)
   \\ \midrule
   Speech separation & WSJ0-MIX & SepFormer & Model:  SepFormer (Encoder, MaskNET, inter-intra Transformers, Decoder)\newline
   Augmentation: yes \newline
   Features: waveform \newline
   Pretraining: no \newline
   Dropout: no \newline
   Normalization: LayerNorm \newline
   Number of epochs: 200 \newline
   Batch size: 1 \newline
   Learning rate: 0.00015 \newline
   LR scheduler: ReduceLROnPlateau \newline
   Optimizer: Adam \newline
   Loss: si-snr with pit-wrapper \newline
   Training Time: 3h 00 min/epoch (NVIDIA V100)
   \\ \midrule
   \end{tabular}
\end{table}

\end{document}